\documentclass[%
 reprint,
 amsmath,amssymb,
 aps,
]{revtex4-2}

\usepackage{graphicx}
\graphicspath{{figures/}}
\usepackage{dcolumn}
\usepackage{bm}
\usepackage{braket}
\usepackage[
range-units = single,
]{siunitx}
\usepackage{xspace}

\usepackage{xcolor}
\usepackage[colorlinks=true, allcolors=blue]{hyperref}

\usepackage{xstring}
\newcommand*{\aref}[1]{%
	\IfBeginWith{#1}{eq:}{Eq.~\eqref{#1}}{}%
	\IfBeginWith{#1}{fig:}{Fig.~\ref{#1}}{}%
	\IfBeginWith{#1}{tab:}{Table~\ref{#1}}{}%
	\IfBeginWith{#1}{appendix:}{Appendix~\ref{#1}}{}%
	\IfBeginWith{#1}{sec:}{Section~\ref{#1}}{}%
}

\newcommand{\ca}{$^{40}\text{Ca}^+$\xspace}
\DeclareSIUnit{\gauss}{G}

\usepackage{subfiles}

\newcommand{\thetitle}{Control of an atomic quadrupole transition in a phase-stable standing wave}

\newcommand{\theauthors}{
\author{Alfredo Ricci Vasquez}
 \email{aricci@ethz.ch}
 \affiliation{Institute for Quantum Electronics, ETH Z\"{u}rich, 8093 Z\"{u}rich, Switzerland}

 \author{Carmelo Mordini}
  \affiliation{Institute for Quantum Electronics, ETH Z\"{u}rich, 8093 Z\"{u}rich, Switzerland}

\author{Chloé Vernière}
  \affiliation{Institute for Quantum Electronics, ETH Z\"{u}rich, 8093 Z\"{u}rich, Switzerland}

\author{Martin Stadler}
  \affiliation{Institute for Quantum Electronics, ETH Z\"{u}rich, 8093 Z\"{u}rich, Switzerland}
  
\author{Maciej Malinowski}
  \affiliation{Institute for Quantum Electronics, ETH Z\"{u}rich, 8093 Z\"{u}rich, Switzerland}
  
\author{Chi Zhang}
  \affiliation{Institute for Quantum Electronics, ETH Z\"{u}rich, 8093 Z\"{u}rich, Switzerland}

\author{Daniel Kienzler}
  \affiliation{Institute for Quantum Electronics, ETH Z\"{u}rich, 8093 Z\"{u}rich, Switzerland}
  
\author{Karan K. Mehta}
  \affiliation{Institute for Quantum Electronics, ETH Z\"{u}rich, 8093 Z\"{u}rich, Switzerland \\ Current address: School of Electrical and Computer Engineering, Cornell University, Ithaca, NY 14853, USA}
  
\author{Jonathan P. Home}
\email{jhome@ethz.ch}
  \affiliation{Institute for Quantum Electronics, ETH Z\"{u}rich, 8093 Z\"{u}rich, Switzerland}
  \affiliation{Quantum Center, ETH Z\"{u}rich, 8093 Z\"{u}rich, Switzerland}

}

\begin{document}


\title{\thetitle}

\date{\today}
\theauthors

\begin{abstract}

Using a single calcium ion confined in a surface-electrode trap, we study the interaction of electric quadrupole transitions with a passively phase-stable optical standing wave field sourced by photonics integrated within the trap. We characterize the optical fields through spatial mapping of the Rabi frequencies of both carrier and motional sideband transitions as well as AC Stark shifts. Our measurements demonstrate the ability to engineer favorable combinations of sideband and carrier Rabi frequency as well as AC Stark shifts for specific tasks in quantum state control and metrology.

\end{abstract}

\maketitle

Light-matter interaction is a topic of fundamental interest, which lies at the heart of our technological capability to control quantum matter. The strongest interactions are due to coupling of the field to the electric dipole moment, which exhibit coupling rates dependent on the electric field strength and polarization. These have diverse applications, including  laser cooling and optical trapping \cite{Phillips85,Grimm95,Eschner03}.
When dipole coupling is forbidden by symmetry, electric quadrupole terms can become dominant. Due to their narrow linewidths, such transitions have found an important role in quantum simulation of interacting systems \cite{Blatt2012}, metrology \cite{Ludlow2015, Ray2020, Tojo2004}, precision measurement \cite{Zhang2020} and quantum computing \cite{Leibfried2003,Bruzewicz2019} with both neutral atoms and atomic ions. Quadrupole transitions are driven by electric field gradients, which means that the matrix elements have a deeper tensorial structure than for dipole transitions, as illustrated in various studies, including those of the interaction of single ions with structured light fields \cite{Mundt2002, Schmiegelow2016a,Drechsler2021}. 

Spatial structuring of light fields, achieved through phase-stable interference, is widely used in atomic physics experiments with neutral atoms \cite{Gross2017,Subhankar2019}. Standing waves may also carry advantages for quantum information processing with trapped ions, where spatial structure can be exploited to control the coupling of an atom by tuning its position inside the field. Driving transitions in a region of zero electric field  suppresses any Stark shifts due to non-resonant dipole couplings. These features can be relevant in the context of achieving faster entangling gates \cite{Mehta2019,Sorensen2000}, for applying spin dependent forces \cite{deNeeves2022}, and for metrology \cite{Huntemann2012}. However, these applications require a phase stable standing wave, which must be positioned precisely relative to the position of the ion. Such control has been achieved with free space laser beams using active feedback stabilization \cite{Schmiegelow2016}, using reflections from the trap surface \cite{Delaubenfels2015}, or using a standing wave produced by a high finesse optical cavity \cite{Mundt2002}. However all of these approaches pose significant challenges for scaling systems up to many laser beams, such as will be required for high performance quantum computation, compact atomic clocks, or sensing.


In this Letter, we demonstrate control of a \ca ion quadrupole transition in a phase stable standing wave \cite{Mehta2016, Niffenegger2020, Mehta2020, Ivory2020} generated through the use of integrated optics: An on-chip waveguide splitter feeds two grating output couplers which emit into free space. The resulting optical field is a standing wave in the direction parallel to the chip plane and a traveling wave in the direction perpendicular to the chip, which causes a spatial variation in the relative strength of the allowed transitions dependent on the ion position inside the electric field pattern. 
We characterize the matrix elements of the allowed transitions of the trapped ion as a function of its position. We separate for each of them the strength of the resonant coupling measured by the Rabi frequency $\Omega$ from the AC Stark shift induced by non-resonant excitations. We explore the relation between carrier transitions and the corresponding sidebands along the axial motional mode aligned with the direction of the standing wave, and find positions with favorable properties, e.g., offering sideband transitions with no accompanying carrier excitation nor AC Stark shift. This provides the basis for exploiting such a light field for quantum computation, choosing the suitable ion position to suppress unwanted off-resonant effects in gates performed on the optical qubit of the ion.

\begin{figure}[ht!]
    \centering
    \includegraphics[width=\columnwidth]{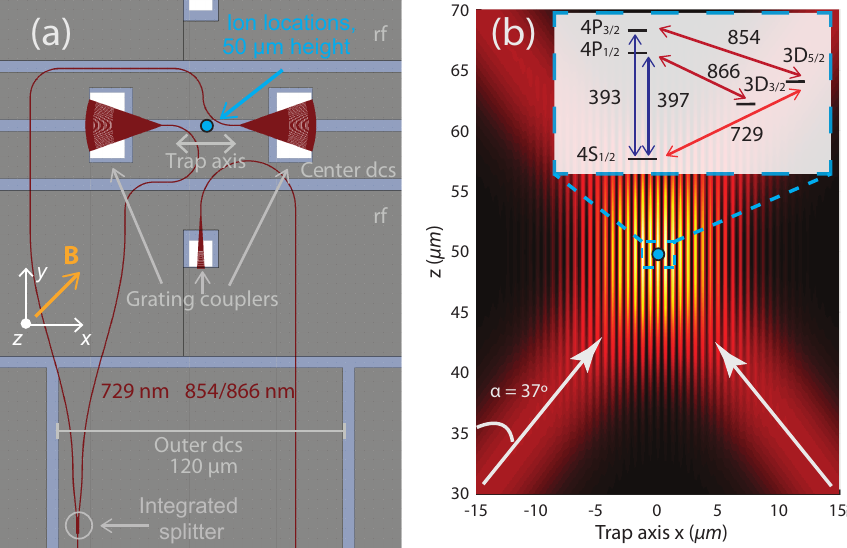}
    \caption{(a) Trap layout schematic. Two grating couplers emit two \SI{729}{\nano\meter} beams generated using an integrated waveguide splitter. The off-axis coupler is used for \SI{854}{\nano\meter} and \SI{866}{\nano\meter} repump light. Near-UV light is sent through free-space. The ion is trapped \SI{50}{\micro\meter} above the position of the green dot. A set of DC electrodes is used to position the ion along the trap axis ($x$) with a resolution on the order of $\sim\SI{10}{\nano\meter}$. (b) Schematic of the intensity profile of the light field along the $xz$ plane near the position of the ion. The inset shows the electronic structure of the \ca ion, where the numbers are the wavelengths of the transitions in nm. Zeeman sublevels are not shown.} 
    \label{fig:setup}
\end{figure}

Figure~\ref{fig:setup} shows the geometry of the trap and the transitions relevant for our experiment. Two grating outcouplers aligned along the trap axis $x$ each emit a Gaussian-like laser beam at \SI{729}{\nano\meter} propagating at an angle $\alpha = \SI{37}{\degree}$ from the normal to the trap plane $z$. The beams intersect at a height of \SI{50}{\micro\meter} above the chip where an ion is trapped, creating the standing wave interference pattern.
Both laser beams are linearly polarized along the $y$ direction, and the resulting field can be described in the vicinity of the ion as a combination of two plane waves
\begin{equation}
\label{eq:e_field}
    \mathbf{E}(\mathbf{r}) = E_0 e^{ik_zz}\cos(k_x \Delta x)\mathbf{e}_y,
\end{equation}
where $k_x = k \sin(\alpha)$ and $k_z = k \cos(\alpha)$ are the components of the beam wavevector, $x$ is the ion position and $k = 2\pi/\lambda $ and $\mathbf{e}_y$ is the unit vector along $y$, and $\Delta x = (x-x_0)$, where $x_0$ accounts for the shift in the standing wave phase relative to the co-ordinate system. 

This field is used to drive quadrupole transitions between our chosen ground state $\ket{4S_{1/2}, m_j=-1/2} = \ket{g}$ and the $3D_{5/2}$ level. Multiple transitions exist from $\ket{g}$, to states in the $3D_{5/2}$ manifold differing in their magnetic quantum number by $\Delta m_j$, each of which are spectrally resolved. For a given component with resonant frequency $\omega_o$, the strongest resonant excitations (carrier transitions) occur when the laser frequency $\omega_l = \omega_o$, resulting in the Hamiltonian $\hat{H}_{\textrm{c}}=(\hbar/2)(\Omega_c \hat\sigma_+ + {\rm h.c.}) $,
where $\hat\sigma_+ = \ket{e}\bra{g}$ and $\hat\sigma_- = \ket{g}\bra{e}$ are the atomic raising and lowering operators, respectively, and $\Omega_c$ is the carrier Rabi frequency, given by
\begin{equation}
    \label{eq:rabi_freq}
    \Omega_c = \frac{eE_0}{\hbar} \mathbf{F}(\Delta m_j, \mathbf{B}) \cdot \pmb{\kappa}_c(x), 
\end{equation}
where $\pmb{\kappa}_c = \{-\sin(\alpha)\sin(k_x\Delta x), 0, i \cos(\alpha)\cos(k_x\Delta x)\}$ encodes the gradient of the electric field as a function of the ion position, and the matrix elements $F_a(\Delta m_j, \mathbf{B}) = (k/2)\bra{e}\hat{r}_{a}\hat{r}_{y}\ket{g}$ depend on the change in the magnetic quantum number in the selected transition $\Delta m_j$, as well as the direction of the external magnetic field $\mathbf{B} = (\mathbf{e}_x + \mathbf{e}_y) / \sqrt 2$ defining the quantization axis.
In this configuration, at the antinodes of the standing wave (i.e., $k_x\Delta x= p \pi$, $p\in \mathbb{Z}$) there are only gradients of the fields in the out-of-plane $z$ direction which are maximised at this position. At the nodes ($k_x\Delta x=\pm(p + 1/2)\pi$), there are only gradients of the fields along the trap axis $x$ direction, which are maximized at this position. Explicit calculation of the matrix elements gives $F_z(\Delta m_j=0) = F_x(\Delta m_j=\pm1) = 0$ such that at the antinodes the carrier transition with $\Delta m_j = 0$ is suppressed while the $\Delta m_j = \pm1$ transitions are maximized. The opposite happens at the nodal positions. Furthermore, the relative phase of $90^\circ$ between $F_z(\Delta m_j=\pm2)$ and $F_x(\Delta m_j=\pm2)$ allows one to tune the coupling strength of these transitions, maximising (and minimising) them in between the nodes and anti-nodes of the standing wave.
Detailed calculations of the matrix elements $F_a$ are provided in the Supplementary Material (SM) 


Since the centre of mass of the ion oscillates in its confining potential, the laser light is phase modulated in the rest frame of the atom and the spectrum of the light-matter interaction exhibits motional sidebands.  Tuning the laser frequency to the blue sideband of a given transition $\omega_l = \omega_o + \omega_x$, where $\omega_x$ denotes the axial trapping frequency, produces the Hamiltonian $\hat{H}_{\textrm{bsb}} = (\hbar/2)(\Omega_s \hat a^\dagger\hat\sigma_+ + \mathrm{h.c.})$. Here $\Omega_s = a_x \partial_x\Omega_c$ is the sideband Rabi frequency which is defined by the spatial gradient of the carrier coupling along the respective oscillation direction, $a_x = \sqrt{\hbar/ 2 m \omega_x}$ is the zero point motion root-mean-square amplitude and $\hat{a}^\dagger$ and $\hat{a}$ are the creation and annihilation operators, respectively, of the oscillator. In the standing wave the electric field gradient and its derivative along the trap axis are out-of-phase, therefore, for any given transition $|\Omega_s|$ is maximized when $|\Omega_c|$ is at a minimum and vice-versa. This means that the logic regarding the transition Rabi frequency at nodes and antinodes given above for the carrier transitions is reversed for the sidebands. Careful choice of the transition allows suppression of unwanted couplings while implementing a desired Hamiltonian.  


We probe the generated light field by placing the ion at different positions along the trap axis and measuring the respective Rabi frequencies. 
Each repetition of the experiment we cool the axial motional mode of the ion near the ground state ($\bar{n}_x\sim1$ quanta) and prepare the electronic state in $\ket{g}$ via optical pumping. We then excite the transition of interest using a fixed duration pulse of the standing wave and subsequently measure the ion electronic state using state-dependent fluorescence. This sequence is repeated multiple times for each experimental setting to gain statistics. Rabi frequencies for carrier transitions are extracted from the time of minimum occupation of $\ket{g}$, with a pre-calibrated correction for finite switching times of the pulse (SM). For sideband transitions, we extract the Rabi frequency from multiple Rabi oscillations assuming a thermal distribution of the excited motional mode (SM). 
We perform experiments at positions separated by $\SI{15.7}{\nano\meter}$ over a full period of the standing wave. For each position, we probe three of the allowed transitions $\ket g \leftrightarrow \ket{3D_{5/2}, m_j = -5/2, -3/2, -1/2}$ that have $\Delta m_j=-2,-1,0$, respectively.
In our magnetic field of $\SI{5.8}{\gauss}$ the carrier transitions are separated by $\sim\SI{9.7}{\mega\hertz}$, and the trap frequency is $\omega_x = (2\pi)\times \SI{1.64}{\mega\hertz}$.

Figure~\ref{fig:results} shows the measured Rabi frequencies for carriers and sidebands, compared with theoretical predictions. We see broad agreement between experiment and theory for both datasets.
We fit \aref{eq:rabi_freq} to the data for $\Delta m_j = 0$ with $E_0$ and $x_0$ floated and the orientation of the magnetic field fixed to $\mathbf B$. The fit is plotted as well as the resulting predictions for the other transitions (solid curves). There are observable discrepancies between these predictions and the data. We found that these can be reduced by adjusting the direction of the magnetic field in the model by $\SI{2}{\degree}-\SI{3}{\degree}$ in both the $x-y$ and the $x-z$ planes (dashed lines). This adjustment is consistent with uncertainties in the magnetic field direction estimated previously in this setup
\cite{Zhang2022}. As expected the minimum values for the excitation of carrier happen when the sideband excitation is maximized. At these positions, the carrier Rabi frequencies are suppressed relative to their maximal values by $14.07(6)\times$, $16.9(1)\times$ and $35.9(4)\times$ for $\Delta m_j = 0,-1,-2$ respectively.  We have observed suppression factors up to $300\times$, but do not find that this is repeatable. This is likely due to slight changes in the exact orientation of the magnetic field, for which the transition with $\Delta m_j = 0$ has the highest sensitivity and the transition with $\Delta m_j = -2$ has the lowest, in agreement with our observations. The measurements of the sideband Rabi frequencies exhibit higher uncertainties than the carrier counterparts given their dependence on the phonon occupation which follows a shot-to-shot thermal distribution. Also, since $|\Omega_s|$ is $\eta = k_x a_x \sim 0.05$ times lower than $|\Omega_c|$, $|\Omega_s|$  is more sensitive to detunings arising from miscalibration in the carrier transition frequency, AC Stark shift and trap frequency.

\begin{figure}
    \includegraphics[width=\columnwidth]{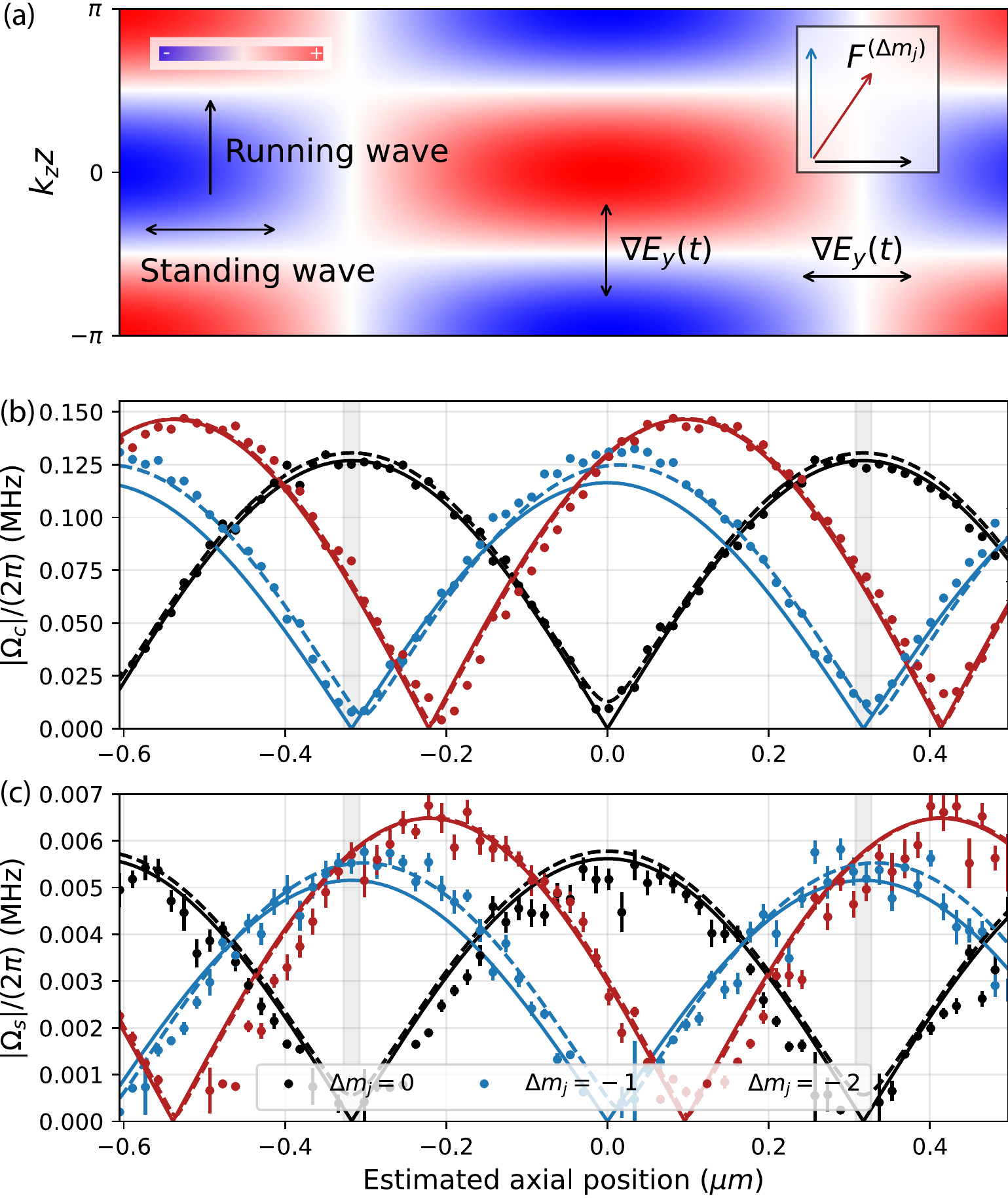}
    \caption{(a) Sketch of the spatial dependence of the y-polarised electric field (Eq.~\ref{eq:e_field}) as a function of $z$ and $x$ at a fixed instant in time. The arrows indicate the direction of the electric field gradient. The box in the top right corner indicates the directions of the magnitude of the components of $\mathbf{F}(\Delta m_j)$ in the $xz$ plane, with colours chosen to match the respective data sets in (b) and (c). (b-c) Measurements of absolute value of the carrier (b) and blue sideband (c) Rabi frequencies as a function of the ion position. Solid lines plot predictions from the theoretical model in Eq.~\ref{eq:rabi_freq} with the nominal orientation of the magnetic field, while dashed lines include a correction for the magnetic field orientation of $2.7^o$ and $3.1^o$ for the out-of- and in- plane directions, respectively. Two shaded lines are used to denote the positions of the nodes of the standing wave.}
    \label{fig:results}
\end{figure}

When resonantly exciting one of the transitions between the $4S_{1/2}$ and $3D_{5/2}$ manifolds, all the other transitions in the ion are driven off-resonantly. This results in a net AC Stark shift of the desired transition \cite{Haffner2003}. The main contribution to the AC Stark shift comes from off-resonantly driving $3D_{5/2} \leftrightarrow 4P_{3/2}$, $4S_{1/2} \leftrightarrow 4P_{1/2}$ and $4S_{1/2} \leftrightarrow 4P_{3/2}$  transitions, which are all dipole-allowed and, therefore, proportional to the intensity of the field, while a second contribution applies from off-resonantly exciting other quadrupole transitions in the $4S_{1/2} \leftrightarrow 3D_{5/2}$ manifold (see \aref{fig:setup}(b)). At the node of the standing wave we can maximally drive the carrier transition with $\Delta m_j=0$ or for the sideband transition with $\Delta m_j = -1$, allowing us to suppress the dipole contribution of the AC Stark shifts. When maximally driving any sideband, the dominant AC Stark shift contribution from off-resonantly driving the carrier is suppressed.
Figure~\ref{fig:stark-shifts} shows the measured AC Stark shifts for the carrier and the sidebands. Around the nodal position where the $\Delta m_j=0$ carrier and the $\Delta m_j=-1$ sideband are strongly driven, we observe a high suppression of the AC Stark shift for both transitions.
The observed residual shift is due to off-resonant couplings to neighboring quadrupole transitions. In practice, this residual quadrupole AC Stark shift could be minimized by increasing the Zeeman splitting with higher magnetic fields. Detailed calculations on the separate contributions of the dipole and quadrupole AC Stark shifts can be found in the SM.

\begin{figure}[ht!]
    \centering
    \includegraphics[width=\columnwidth]{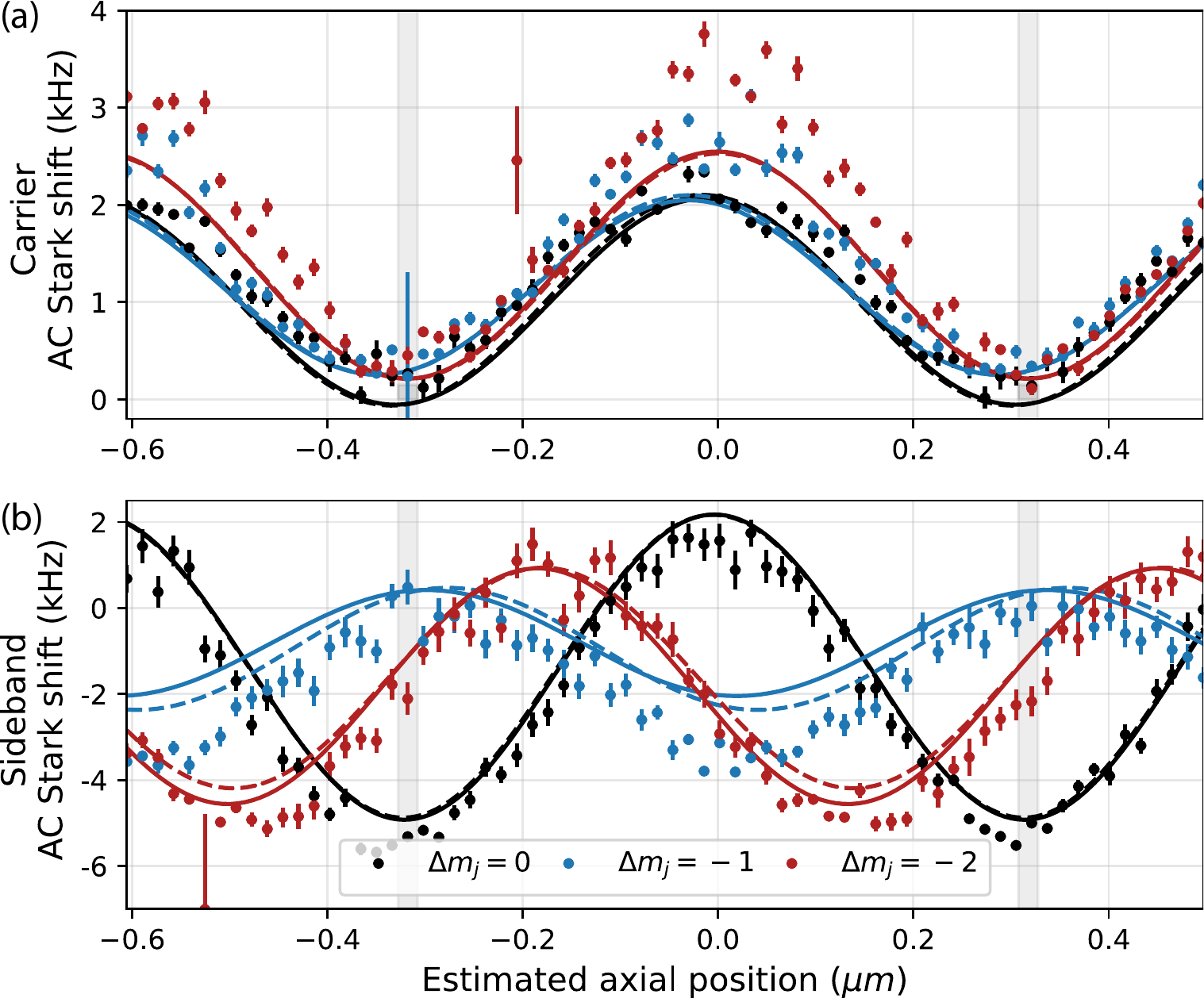}
    \caption{Measurements of the AC Stark shifts on the carrier (a) and on the sideband transitions (b) as a function of the ion position in a period of the standing wave. Solid lines are computed using the nominal magnetic field orientation and the amplitude of the light field obtained from the Rabi frequency pattern of the transition with $\Delta m_j=0$. Dashed lines include a correction for the magnetic field orientation.}
    \label{fig:stark-shifts}
\end{figure}

The use of integrated photonics provides a reliable way of passively controlling the relative phase of the beams generating the standing wave. The degree with which the ion can be placed with respect to the standing wave is limited by stray electric fields. 
We studied the stability of this positioning by repeatedly recording the carrier Rabi frequency pattern and measuring the displacement of the ion with respect to the standing wave pattern during ~10h. The primary cause of shifts are due to ultra-violet light at 389~nm and 423~nm used for loading ions by photo-ionization. This is illustrated in figure \ref{fig:repetition}(a), which shows the fitted value of $x_0$. Between each of the first 5 measurements we turn on the photoionisation (PI) beams for a 5 minute duration. On both trials, we see a similar drift while the PI light is on, and residual drift at the level of $\sim \SI{10}{\nano\meter}$ when it is off. The position displacement tends to saturate after a few cycles of exposure to PI light, and is then followed by a discharge process that occurs within the first hours but leaves a permanent displacement over longer time scales. This behavior was repeatable over several trials. The dependence of the fitted $x_0$ on the presence of PI light suggests that the origin of the displacement is dominated by drifts in the ion position rather than shifts in the relative phase between the beams forming the standing wave \cite{Harlander2010}.

At timescales faster than those required to obtain a single Rabi frequency, changes in the ion position would produce decay in the observed Rabi oscillations. We measured the decay of the Rabi oscillations as a function of the position of the ion in the standing wave for the carrier transition with $\Delta m_j= 0$. Assuming shot-to-shot fluctuations of the Rabi frequency sampled from a Gaussian distribution with width $\sigma_\Omega$, the population of the $\ket{S_{1/2}}$ state as a function of time is found to be $P(\ket{S_{1/2}}) = 0.5 +0.5\exp{(-\sigma^2_\Omega t
^2/2)}\cos(|\Omega|t)$. Extracted values of $\sigma_\Omega/|\Omega|$ as a function of the position of the ion  are shown in figure \ref{fig:repetition} (b), exhibiting increased Rabi frequency fluctuations around $k_x \Delta x = 0$. We fit these results with two different models, where fluctuations of the Rabi frequency are produced  by fluctuations of (i) the out-of-plane direction of the magnetic field or (ii)  small displacements between the ion and the light field. Both models produce satisfactory fits, allowing us to bound the shot-to-shot magnetic field fluctuations to $\sigma_\mathbf{B} = 0.25^o$ or, alternatively the position fluctuations to $\sigma_x = 1.6$ nm, the latter representing a fluctuation of $\sim 0.13\%$ of the period of the standing wave, either arising from changes of the relative phase between the two beams or drifts in the ion position. The offset to the curve due to Rabi frequency fluctuations on the order of $\sigma_\Omega/|\Omega| \sim 1.5\% $ is consistent with the expected thermal occupation of the atomic center of mass motion \cite{Malinowski2021}. Thermal effects in the carrier Rabi frequency can be accounted by considering higher order terms in the Lamb-Dicke expansion, which modify the carrier Rabi frequency to $\Omega_c(n) \approx \Omega_c(1-\eta^2(2n+1)/2)$ for a given Fock state $n$. Since the fluctuations are proportional to the Rabi frequency the fractional fluctuations are independent of position.

\begin{figure}[b!]
    \includegraphics[width=\columnwidth]{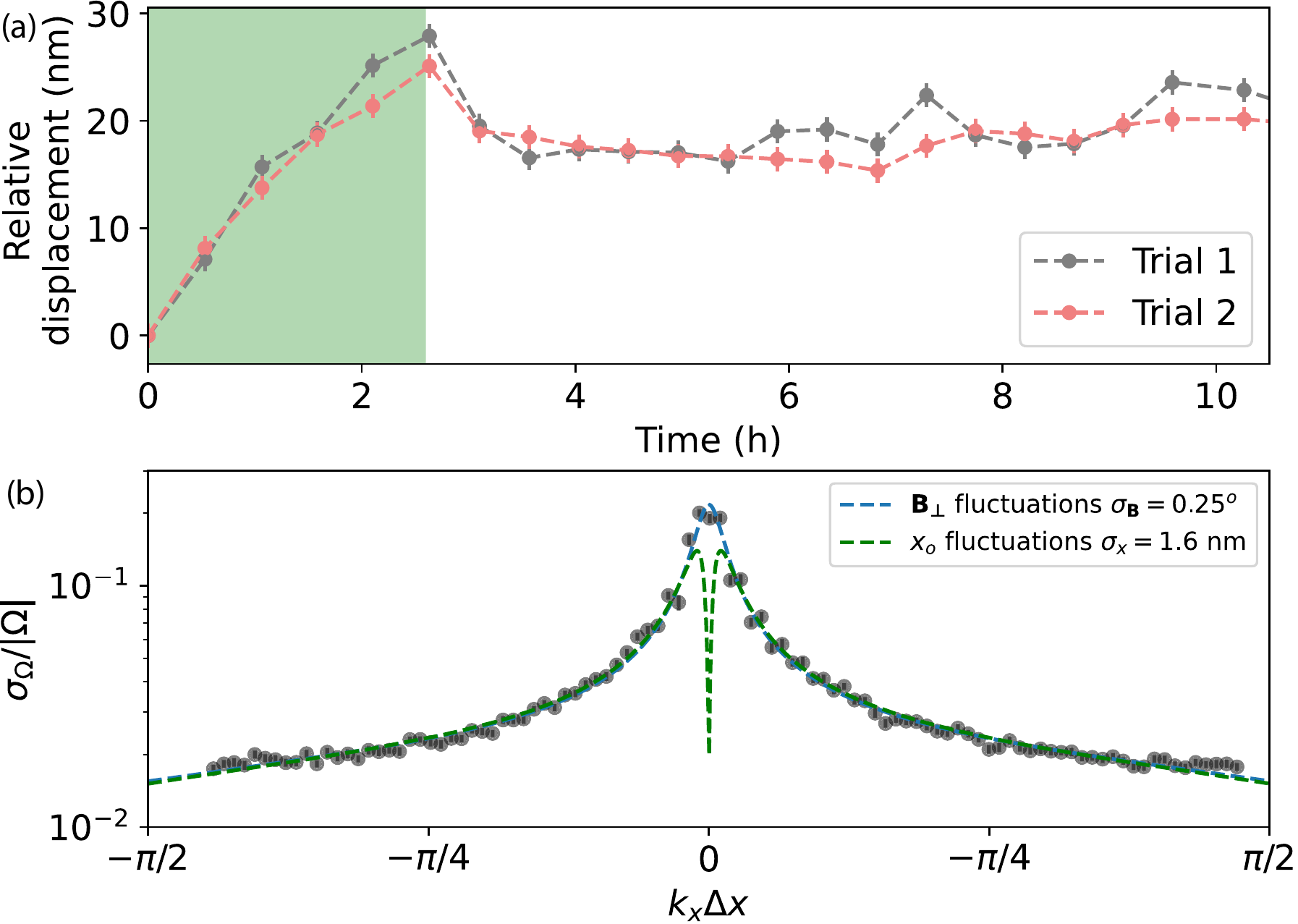}
    \caption{
    (a) shows
    the relative displacement of the ion position with respect to the standing wave pattern over time ($x_0)$. Two repetitions of the same experiment performed in different days are shown. The green region shows the time span where 5 minutes of exposure to PI light was interleaved between measurements. (b) shows the relative standard deviation of the $\Delta m_j=0$ Rabi frequency as a function of the position of the ion in the standing wave. Lines show two different fits, corresponding to Rabi frequency fluctuation coming from position or magnetic field orientation fluctuations.}
    \label{fig:repetition}
\end{figure}

Our work offers insight in the physics of atom-light interaction at a fundamental level, and demonstrates the use of integrated photonics to create structured light fields with chosen properties that can be exploited advantageously for quantum computing  and metrology. Specifically, the possibility of precisely locating an ion in the standing wave without constant re-calibration or active feedback allows us to envision the use of such a device in large scale quantum information processors. In the specific field configuration employed here, the choice of the $\Delta m_j=-1$ transitions maximizes sideband Rabi frequency at intensity nulls, where AC Stark shifts are minimal and the carrier transition is suppressed. For quantum gates driven by motional sidebands, this offers a route to reduced off-resonant contributions to gate errors. The carrier transition with lowest magnetic field sensitivity ($\Delta m_j=0$) is driven with maximal strength at the nodes of the standing wave, facilitating use for optical clocks with reduced sensitivity to AC Stark shifts. All of these features are available in an integrated form, which greatly facilitates scaling and portability.\\

We acknowledge funding from the Swiss National Science Foundation (Grant No. PZ00P2\_179909), the EU H2020 FET Open project PIEDMONS (Grant No. 801285) the National Centre of Competence in Research for Quantum Science and Technology (QSIT) and the EU Quantum Flagship H2020- FETFLAG-2018-03 under Grant Agreement No. 820495 AQTION. KM acknowledges support from an ETH Post-doctoral fellowship. We acknowledge LioniX International for fabrication of the trap devices. 

KM conceived the experiment and designed the device. Experimental data were taken by AR and CM, using an apparatus with significant contributions from MM, CZ, MS, and KM. Data analysis and modeling were performed by AR, CV and CM. The paper was written by AR and CM with input from all authors. The work was supervised by DK and JH.



\bibliography{bibliography}

\newpage
\clearpage
\pagebreak
\renewcommand{\bibliography}[1]{}  

\setcounter{equation}{0}
\setcounter{figure}{0}
\setcounter{table}{0}
\setcounter{footnote}{0}
\setcounter{page}{1}
\makeatletter
\renewcommand{\theequation}{S\arabic{equation}}
\renewcommand{\thefigure}{S\arabic{figure}}
\renewcommand{\bibnumfmt}[1]{[S#1]}
\makeatother

\title{Supplemental materials for:\\\thetitle}
\theauthors 

\maketitle

\section{Calculation of Rabi frequencies}
\label{ap:rabis}

For an interaction Hamiltonian given by the quadrupole term of the multipole expansion, the Rabi frequencies between a ground state $\ket{g}$ and an excited state $\ket{e}$ of the single valence electron of an ion are are given by \cite{Jackson1962, Leibfried2003}

\begin{equation}
    \Omega_c = \sum_{i,j}\frac{e}{2\hbar}\bra{g} \hat r_{i} \hat r_{j}\ket{e}\frac{\partial E_j}{\partial r_i},
\end{equation}

where $\hat{r}$ is the position of the valence electron of the ion, and the indices $i$ and $j$ are for the Cartesian coordinates. If we consider an electric field where the spatial structure is given by,

\begin{equation}
    \mathbf{E(r})=E_0e^{ik_zz}\cos(k_x\Delta x)\mathbf{e}_y,
\end{equation}

it is then possible to write the gradient of the field as

\begin{equation}
    \partial_i E_y=E_0\left<-k_x\sin(k_x\Delta x),0,ik_z\cos(k_x\Delta x)\right>,
\end{equation}

where the term $e^{ik_zz}$ is factored as a global phase. Then, for a beam emitted at an angle $\alpha$ from the vertical such that $k_x=k\sin(\alpha)=\frac{2\pi}{\lambda}\sin(\alpha)$ and $k_z=k\cos(\alpha)=\frac{2\pi}{\lambda}\cos(\alpha)$, one gets $\partial_i E_y=E_0k\kappa_i$ with 

\begin{equation}
    \pmb{\kappa}_c= \left<-\sin(\alpha)\sin(k_x\Delta x),0,i\cos(\alpha)\cos(k_x\Delta x)\right>.
\end{equation}

Defining $F_i(\Delta m_j,\mathbf{B}) = (k/2)\bra{g}\hat r_{i}\hat r_{y}\ket{e}$ leads to Rabi frequencies definitions from the main text,

\begin{equation}
\Omega_c = \frac{eE_0k}{2\hbar}\sum_{i}\bra{e}\hat r_{i}\hat r_{y}\ket{g}\mathbf\kappa_i,
\end{equation}

where experimentally we can only access the magnitude of the Rabi frequency $|\Omega_c|$ given by

\begin{equation}
    |\Omega_c| = \frac{eE_0}{\hbar}\left|\mathbf F(\Delta m_j, \mathbf{B})\cdot\pmb\kappa_c\right|.
\end{equation}

Now, it is possible to write the matrix elements $\bra{e} \hat r_{i} \hat r_{j}\ket{g}$ as,

\begin{equation}
    \bra{e}\hat r_{i}\hat r_{j}\ket{g}=\bra{g}r^2C^{(2)}\ket{e}\sum_{q=-2}^2\begin{pmatrix}j&2&j'\\-m_j&q&m_j'\end{pmatrix}(R^Tc^{(q)}R)_{ij}.
\end{equation}

where $\bra{e}r^2C^{(2)}\ket{g}$ are the \textit{reduced matrix elements}, $c^{(q)}$ are the rank-2 spherical basis tensors and the quantity in parenthesis are \textit{Wigner 3-j numbers} \cite{James1998}. $R$ is the matrix rotating the coordinate system of the ion (where z is the quantization axis), to the trap coordinate system as specified in the main text. For a given magnetic field orientation $\mathbf{B}=\left<b_x,b_y,b_z\right>$, $R$ can be expressed as

\begin{equation}
\label{eq:rotation}
    R = I_{3\times3}-\left[\begin{matrix} \frac{b_{x}^{2}}{b_{z} + 1}  & - \frac{b_{x} b_{y}}{b_{z} + 1} & - b_{x}\\- \frac{b_{x} b_{y}}{b_{z} + 1} &  \frac{b_{y}^{2}}{b_{z} + 1} & - b_{y}\\b_{x} & b_{y} &   \frac{(b_{x}^{2} + b_{y}^{2})}{b_{z} + 1}\end{matrix}\right],
\end{equation}

which for our experimental nominal orientation $\mathbf{B}=\frac{1}{\sqrt{2}}\left<1,1,0\right>$ reduces to,

\begin{equation}
    R = \frac{1}{2}\left[\begin{matrix}1& -1 & - \sqrt{2}\\-1 & 1 & - \sqrt{2}\\\sqrt{2} & \sqrt{2} & 0\end{matrix}\right].
\end{equation}

To study the effect of misalignment on the magnetic field orientations, we simply replace $\mathbf{B}$ in equation \ref{eq:rotation} with an arbitrary direction.

It is possible to relate the reduced matrix elements to the decay rates $A$ of the transition \cite{James1998, Barton2000} so that,

\begin{equation}
    \left|\bra{g}r^2C^{(2)}\ket{e}\right|^2 = \frac{90A}{c\alpha k^5},
\end{equation}

where $c$ the speed of light and $\alpha$ the fine-structure constant. Considering the initial state to be $\ket{g} = \ket{4S_{1/2}, m_j=-1/2}$ the elements $F_i$ are then given by

\begin{equation}
    F_i=\\
    \beta\begin{pmatrix}1/2&2&5/2\\1/2&-\Delta m_j&\Delta m_j-1/2\end{pmatrix}(R^Tc^{{(-\Delta m_j)}}R)_{i,y},
\end{equation}

with,
\begin{equation}
    \beta = \sqrt{\frac{45A}{2k^3c\alpha}}.
\end{equation}

Explicit calculation of the $F_i(\Delta m_j)$ elements gives,

\begin{equation}
    \mathbf{F}(\Delta m_j) = \beta
    \begin{cases}
    \frac{-\sqrt{10}}{10}\left<\frac{1}{2}, \frac{1}{6},0\right> \text{             if      }  \Delta m_j=0\\
    \frac{-\sqrt{15}}{30}\left<0,\sqrt{2}(1+i), 1-i\right> \text{     if     } \Delta m_j=-1\\
    \frac{-\sqrt{30}}{60}\left<i,-i, -\sqrt{2}\right> \text{      if     } \Delta m_j=-2\\
    \end{cases}
\end{equation}

\section{AC Stark-shifts}

In this appendix we study the different contributions of the AC Stark shifts for the three electronic carriers and their sidebands as a function of the position of the ion within the standing wave. We study two main contribution to the AC Stark shifts, given by off-resonantly driving the dipole transitions in the ion ($\Delta\omega_{dp}$), and off-resonantly driving the neighbour quadrupole transitions ($\Delta \omega_{qp}$) \cite{Haffner2003}. The total frequency shift is then given by,

\begin{equation}
    \Delta\omega = \Delta\omega_{dp}+\Delta\omega_{qp}.
\end{equation}

In order to calculate the dipole Stark shifts, we consider the contribution from off-resonantly driving the $4S_{1/2}\leftrightarrow 4P_{1/2}$ at 397 nm, $4S_{1/2}\leftrightarrow 4P_{3/2}$ at 393 nm and $3D_{5/2}\leftrightarrow 4P_{3/2}$ at 854 nm transitions. When shining light at 729 nm, the transitions at 393 nm and 397 nm are driven red detuned, so the $\ket{4S_{1/2}, -1/2}$ state is \textit{downshifted} and the transition at 854 nm is driven blue detuned so the states in the $3D_{5/2}$ level are \textit{upshifted}. Furthermore, the Stark-shift from the $3D_{5/2}\leftrightarrow 4P_{3/2}$ is dependent on the magnetic number $m_j'$ of the state in the $3D_{5/2} $. The dipole Stark-shift is then,

\begin{equation}
    \Delta\omega_{dp} = \Delta\omega_{S} - \Delta\omega_{D}(\Delta m_j),
\end{equation}

with $\Delta \omega_S$ and $\Delta \omega_D$ the shifts on the $\ket{g}$ and $\ket{e}$ states, respectively. 
Under the assumption of a weak field largely detuned from any dipole transition $\Delta\omega_{dp}$ can be expressed in terms of the local electric field amplitude $|\mathbf{E}|^2$, the frequency of the laser $\omega$ and $\alpha(\Delta m_j) = (\alpha_S(\omega)-\alpha_D(\omega,\Delta m_j))$ with $\alpha_S$ and $\alpha_D$ the \textit{dynamic polarisabilities} of the $\ket{g}$ and $\ket{e}$ states as \cite{Delone1999},

\begin{equation}
    \frac{\Delta\omega_{dp}}{2\pi} = \frac{\alpha( \Delta m_j)|\mathbf E|^2}{4\hbar}. 
\end{equation}

We performed the calculations of the dynamic polarisabilites by directly computing the dipole matrix elements following Ref. \cite{James1998} and verified them independently using the python library \texttt{atomphys} \cite{Grau}. We found $\alpha(\Delta m_j)/\hbar=4.35\times10^{-6}$, $4.13\times10^{-6}$ and $3.68\times10^{-6}$    \SI{}{m^2\hertz /V^2}, for $\Delta m_j=0$, $-1$ and $-2$, respectively.


For the calculation of the quadrupole AC Stark shift $\Delta\omega_{qp}$ we need to independently consider the case when the carrier is driven and when the blue sideband is driven since their detunings to the quadrupole transitions are different. We don't make this distinction in the dipole Stark-shift since the dipole transitions are very far off-resonant compared to the trap frequency. The quadrupole AC Stark-shift when driving the carriers is given by,

\begin{equation}
    \Delta\omega_{qp} (\omega=\omega_0) = \frac{1}{4}\sum_i\frac{\Omega^2_i}{(\omega_i -\omega_0)},
\end{equation}

where $\omega_0$ is the frequency of the transition driven resonantly and the index $i$ runs over the transitions that share a common state with the driven transition. For example if we resonantly drive the $\ket{g}\leftrightarrow \ket{3D_{5/2},-1/2}$ transition we need to consider the Stark-shift from the $\ket{4S_{1/2},-1/2}\leftrightarrow \ket{3D_{5/2},-5/2,-3/2,+1/2+3/2}$ and $\ket{4S_{1/2},+1/2}\leftrightarrow \ket{3D_{5/2},-1/2}$ transitions. $\Omega_i$ is the Rabi frequency of the $i$-th transition, which can be computed following appendix \ref{ap:rabis} and the detuning $\delta_{i} = (\omega_i-\omega_0)$ corresponds to the Zeeman splitting between the different sub-levels. For our magnetic field, the splitting on the $\ket{4S_{1/2}}$ sublevels is $\sim (2\pi)\times\SI{16.2}{\mega\hertz}$ and for the $\ket{3D_{5/2}}$ levels is $\sim (2\pi)\times\SI{9.75}{\mega\hertz}$.

When driving the axial blue sideband (with frequency $\omega_x \sim (2\pi)\times1.6\text{ MHz}$) the carrier transition is now driven off-resonantly leading to a second contribution where the two energy levels of the carrier are shifted and therefore the Stark-shift is given by,

\begin{equation}
    \Delta\omega_{qp}(\omega=\omega_0+\omega_x) = \frac{1}{4}\sum_i\frac{\Omega^2_i}{(\delta_{i}-\omega_x)}-\frac{\Omega_0^2}{2\omega_x}.
\end{equation}

\begin{figure}[ht!]
    \centering
    \includegraphics[width=\columnwidth]{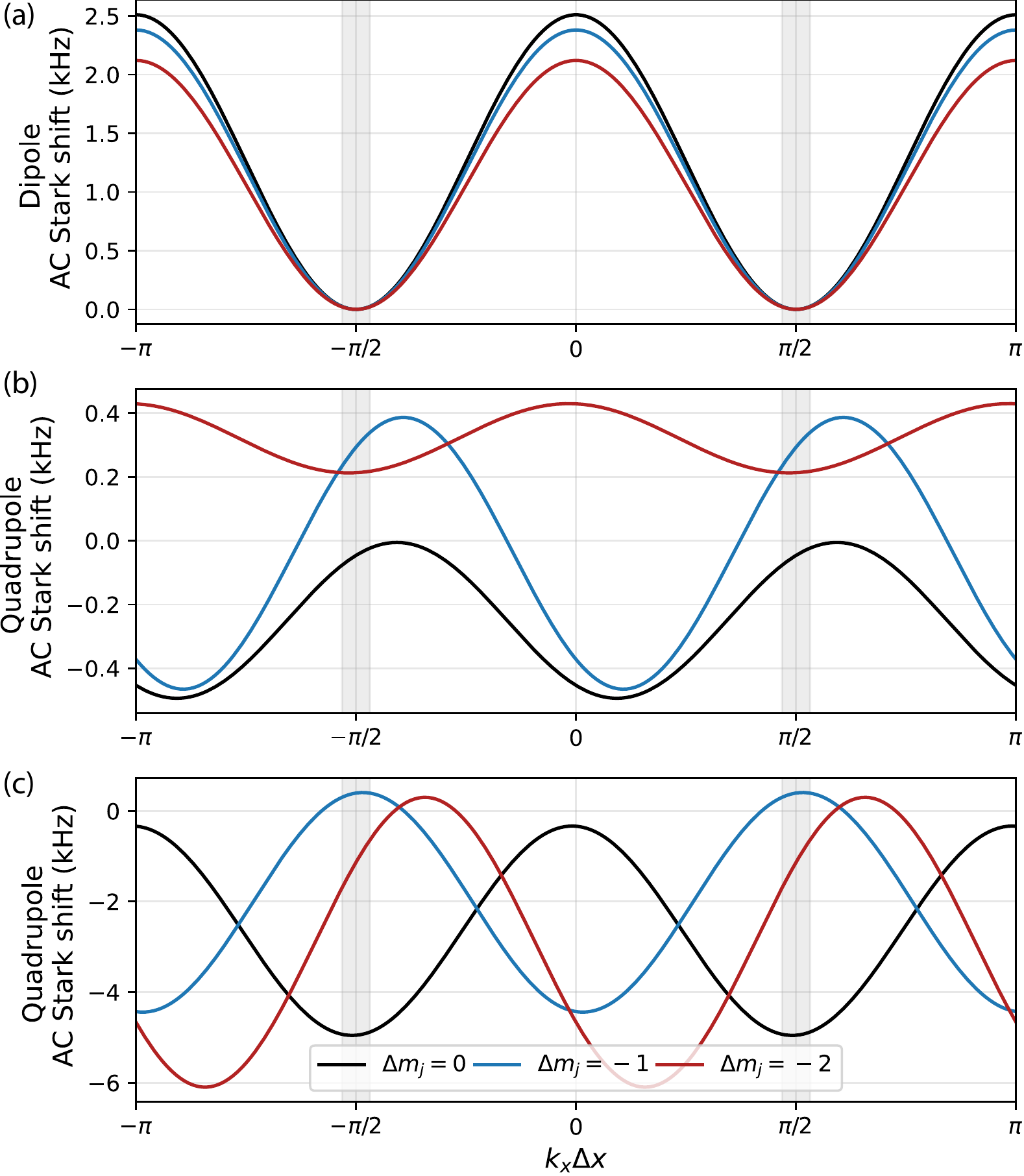}
    \caption{\label{fig:stark-shift-decomposition} Decomposition of the total Stark shifts presented in figure 3(a-b) of the main text into its dipole (a), quadruple components for the carrier (b) and quadrupole components for the sidebands (c).}
\end{figure}

We show the decomposition of the theoretical Stark-shift presented in figure 3 of the main text in its dipole and quadrupole components as a function of the axial position in figure \ref{fig:stark-shift-decomposition}.





\section{Intensity profile of the beam}

\begin{figure}[h!]
    \centering
    \includegraphics[width=\columnwidth]{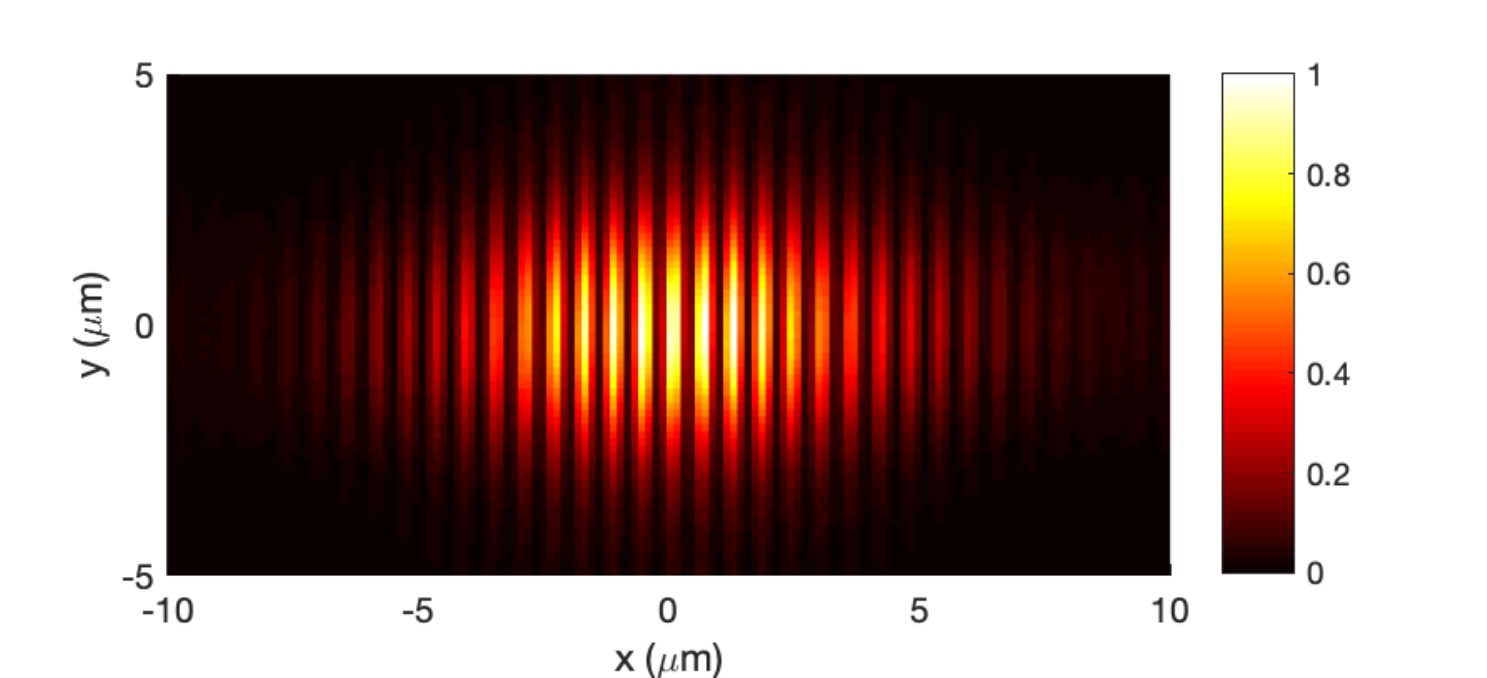}
    \caption{Measurement of the intensity profile of the standing wave in the $xy$ plane. The scale is in normalised arbitrary units.}
    \label{fig:intensity-profile}
\end{figure}

So far in the theoretical model we considered the light emitted from the outcouplers as plane waves, however, in reality the emitted beams are tightly focused, with a Gaussian-like spatial distribution \cite{Mehta2020}. Figure \ref{fig:intensity-profile} shows the measured intensity profile of the light field on the $xy$ plane, measured by imaging the field at $\sim \SI{50}{\micro\meter}$ from the trap surface onto a CCD image sensor through a high-NA microscope objective. We choose to perform the experiments near the origin to maximize the validity of the model of the electric field given by equation 1 of the main text.

\section{Experimental methods}
\label{ap:methods}
\subsection{Experimental apparatus}

In order to minimise the electric noise in the trap electrodes and achieve high vacuum, the trap is placed in a \SI{6}{\kelvin} cryostat, inside a vacuum chamber, where \SI{389}{\nano\meter}, \SI{397}{\nano\meter} and \SI{423}{\nano\meter} light is delivered by free-space optics while \SI{729}{\nano\meter}, \SI{854}{\nano\meter} and \SI{866}{\nano\meter} is delivered through optical fibers into the trap integrated waveguides (see figure \ref{fig:apparatus}). \SI{729}{\nano\meter} light used for coherent control of the qubit is taken from the transmission of a high-finesse cavity and then amplified using an injected diode and a tapered amplifier. Finally a double pass free-space AOM and a single pass fiber AOM are used to deliver light resonant with a specific transition into the ion. 

Light generating the SW is input via a single fiber feeding a single waveguide on the device, and routed to the two opposing gratings in Fig. 1(a) by means of an integrated 50:50 waveguide splitter on chip (labeled in Fig. 1(a)). The short propagation distances between this element and the gratings, combined with the rigid optical paths, ensure a high degree of relative passive phase stability. The splitter used in the current device is a symmetric Y-splitter, modified for fabricable minimum feature sizes \cite{zhang2013compact}. The present design was not fully optimized and excess losses simulated by full 3D finite-difference-time-domain (FDTD) modeling for the splitter as designed are $0.8$ dB. Optimized multi-mode interference (MMI) splitter devices \cite{soldano1995optical} for the same wavelength for a next generation of devices have been designed with simulated excess losses of ${<}0.15$ dB using the same waveguide cross section as in this work. Due to the symmetry of the integrated splitter structures used in this work, in the absence of fabrication imperfections the power splitting ratio is exactly unity. Fabrication imperfections will cause any deviations in splitting ratio. Uneven splitting ratios and differential losses between the two arms will result in the two beams forming the standing wave having different electric field amplitudes. Considering the amplitudes of the two beams to be $E_1$ and $E_2$, and the differential amplitude between the beams $\Delta E= E_2-E_1$, such that for $\Delta E = 0$ we get $E_1=E_2=E_0$, we can express the modified field produced by uneven electric field amplitudes as,

\begin{equation}
    E'_y(\mathbf{r}) = \Delta Ee^{i(k_x\Delta x+k_zz)} + E_1 e^{k_zz}\cos(k_x\Delta x).
\end{equation}

This field has an additional running wave component along the axis of the trap. The gradient of this field results in

\begin{equation}
\label{eq:mod-gradient}
    \partial_i E'_y \propto \Delta E e^{ik_x\Delta x}\left<i\sin(\alpha),0,i\cos(\alpha)\right> + E_1\pmb{\kappa},
\end{equation}

where the first term can limit the achieved extinction ratios. Assuming this imperfect splitting is the only nonideality in the optical and $\mathbf{B}$ fields, comparing the ratio of the measured Rabi frequency of the transition with $\Delta m_j = 0$ at the nodes ($|\Omega_{\text{max}}|$) and antinodes ($|\Omega_{\text{min}}|$) to the ratios of $\partial_xE'_y$ at the nodes and antinodes gives,

\begin{equation}
    \frac{|\Omega_{\text{min}}|}{|\Omega_{\text{max}}|} = \frac{\Delta E}{E_1+\Delta E} = \frac{\Delta E}{E_2}.
\end{equation}

We can relate the electric field amplitude to the optical power in each arm such that

\begin{equation}
    \frac{\Delta E}{E_2} = \frac{\sqrt{P_2}-\sqrt{P_1}}{\sqrt{P_2}},
\end{equation}

where $P_1$ and $P_2$ are the powers for the two beams. We define a power splitting ratio $\gamma$ (equal to 0.5 for an ideal splitter) such that for power $P_i$ input to the splitter, $P_2 = \gamma P_i$ and $P_1 = (1-\gamma) P_i$. Using our experimentally measured suppression factor on the $\Delta m_j=0$ transition, we can write

\begin{equation}
        \frac{|\Omega_{\text{min}}|}{|\Omega_{\text{max}}|} = \frac{\sqrt{\gamma} - \sqrt{1-\gamma}}{\sqrt{\gamma}}=0.071,
\end{equation}

allowing us to bound the splitting mismatch (including possible differential loss in the two waveguide paths following the splitter and gratings) to $|0.5-\gamma|< 0.037$. In reality we believe this imbalance to be significantly smaller since the carrier extinction ratios are highly sensitive to magnetic field direction as well.  

Further details regarding the setup and the integrated photonics can be found at refs  \cite{Malinowski2021,Zhang2022,Mehta2020}.

\subsection{Ion trapping and positioning}

\begin{figure}
    \centering
    \includegraphics[width=\columnwidth]{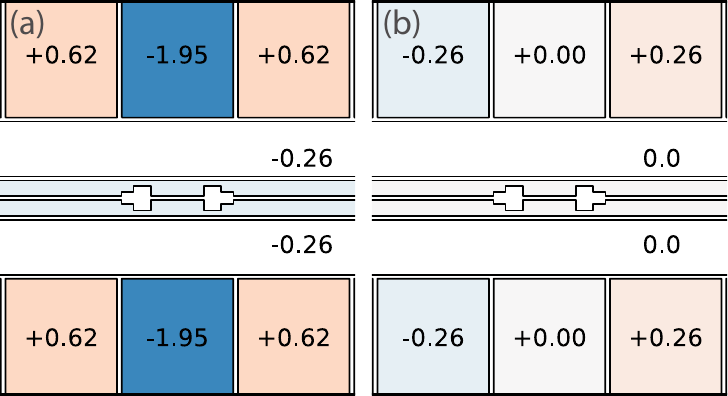}
    \caption{Voltage sets used for trapping (a) and axially positioning the ion (b) plotted on an abstraction of the trap layout. Displayed values are in Volts. The trapping voltage basis produces an axial frequency of \SI{1}{MHz}, and the positioning basis displaces the ion \SI{10}{\micro\meter} in the axial direction for a trap frequency of \SI{1}{\mega\hertz}.}
    \label{fig:positioning}
\end{figure}

In the working zone we manipulate 8 DC electrodes to axially confine and control the position of the ion in the three axes. The basis voltages for confining and axially displacing the ion are shown in figure \ref{fig:positioning}. The displacing voltage set creates a $\sim\SI{153}{\volt/\meter}$ axial electric field at the ion position, which displaces the ion by $\sim\SI{10}{\micro\meter}$ for a trap frequency of 1 MHz. We re-scale the voltage sets to achieve the desired axial trap frequency and displacement of the ion.  Similar voltage sets are used for tilting the radial plane and for minimizing micromotion of the ion.

\begin{figure}

\end{figure}

\begin{figure*}[t!]
\includegraphics[width=\textwidth]{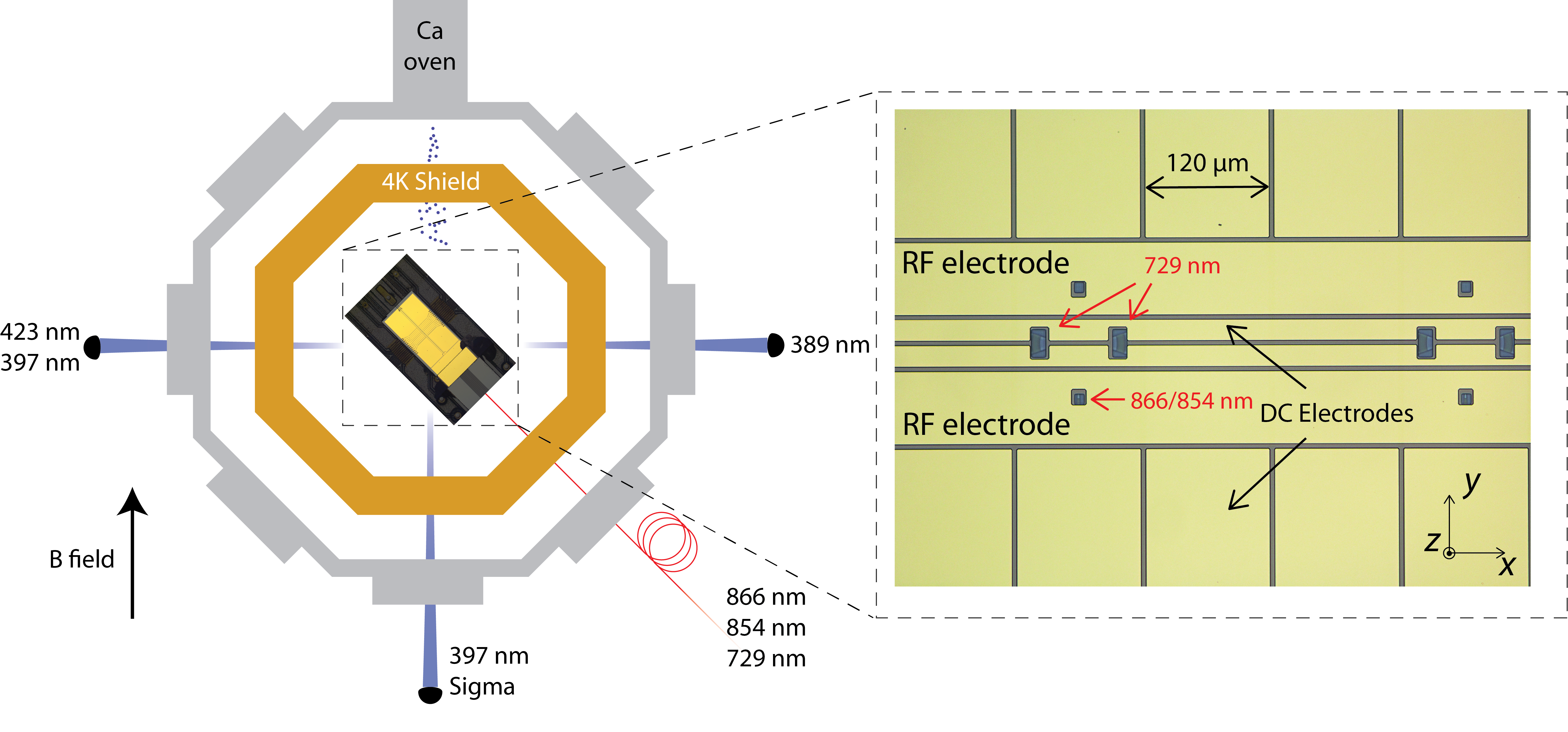}
\caption{\label{fig:apparatus} Schematic showing the main components of our system. The trap lies in a \SI{6}{\kelvin} cryostat used to cool the trap and achieve cryopumping. The \SI{5.8} G magnetic field is generated using permanent magnet disks outside the vacuum chamber. Three viewports are used to shine the ion with UV light used for photoionisation, cooling, state preparation and detection while red and infrared light is delivered via optical fibers directly into the trap. The trap has three experimental zones, each including a waveguide splitter and two outcouplers for \SI{729}{\nano\meter} light as well as an 866/854 \SI{}{\nano\meter} outcoupler. For the experiments we only used one of the experimental zones.}
\end{figure*}

\subsection{Experimental sequence}

For each position and for each of the three transitions we perform the following operations:

\begin{enumerate}
    \item Calibrate the bare frequency of the transition.
    \item Measure the Stark shift on the carrier (Fig 3(a)).
    \item Perform Rabi oscillations of the carrier in resonance with the dressed frequency (bare transition frequency + Stark shift) and extract the carrier Rabi frequency (Fig 2(b)).
    \item Calibrate the trap frequency.
    \item Measure the Stark shift on the sideband (Fig 3(b)).
    \item Perform Rabi oscillations of the sideband ((bare transition frequency + trap frequency + Stark shift)) and extract the sideband Rabi frequency (Fig 3(c)). 
\end{enumerate}

For the six steps we work in two \SI{729}{\nano\meter} optical power regimes. Steps two, three, five and six are all performed at the same high power. Steps one and four are performed at low power.

The bare frequency of the transition is calibrated by performing spectroscopy with a low power pulse of \SI{729}{\nano\meter} light for $\sim\SI{1}{\milli\second}$, such that the Stark shift is negligible at this stage.  

Then we measure the Stark shift on the carrier. We start by applying a pulse at high power resonant with the bare frequency. Since the ion will now experience Stark shifts, this pulse will induce off-resonant Rabi oscillations. From this oscillation we extract the off-resonant $\pi$ time. Then we apply an even amount  of $\pi$ pulses where each of the pulses has a phase difference of \SI{180}{\degree} with the previous one. With this sequence the ion will return to the initial state $\ket{g}$
 when the detuning between the laser and the dressed frequency (bare frequency + Stark shift) is zero. We apply the sequence of pulses for a total time of \SI{150}{\micro\second} and measure the frequency for which the ion returns to the initial state and register the difference between this frequency and the bare frequency as the Stark shift on the carrier.



We proceed to record Rabi oscillations between $\ket{g} \leftrightarrow \ket{e}$ for around one period, where the light at \SI{729}{\nano\meter} is set to be resonant with the dressed frequency of the transition. We extract the first time for which the population transfers from $\ket{g} \rightarrow \ket{e}$ as the $\pi$ time $t_{\pi}$ using a quadratic fit. From $t_{\pi}$ we compute the carrier Rabi frequency as $|\Omega_c|/2\pi = 1/(2t_{\pi} - \Delta t) \approx 0.5/t_{\pi} + \Delta t\left(0.5/t_{\pi}\right)^2$. $\Delta t = \SI{0.335}{\micro\second}$ accounts for the finite switching time of the laser pulse, and is calibrated independently by performing Rabi oscillations and comparing the period of the oscillations, obtained from a sinusoidal fit to the data, to the time of the first maximum inversion of the population fitted using a quadratic fit. 

The blue sideband Rabi frequency and AC Stark shift are measured in a similar way. We perform low power spectroscopy using a ~\SI{500}{\micro\second} pulse around the blue axial sideband frequency followed by one at high power. Here we record the bare blue sideband and dressed blue sideband frequencies and take their differences as the AC Stark shift on the sideband. The Rabi frequency of the sidebands is measured by applying \SI{729}{\nano\meter} light on resonance with the dressed blue sidebands, and then fitting the resulting signal with a statistical mixture of Rabi oscillations weighted over a thermal distribution \cite{Meekhof1996}. 

For each data point on any of the previous step we first apply Doppler cooling to all three motional modes of the ion, then further cool the axial mode using electromagnetically induced transparency (EIT) cooling \cite{Lechner2016}, and optically pump the ion into the initial state using a free-space sigma-polarized \SI{397}{\nano\meter} light.
Then we apply \SI{729}{\nano\meter} light to the ion for the required evolution time, and finally perform state detection by collecting \SI{397}{\nano\meter} fluorescence into a photo-multiplier tube. Integrated light at \SI{866}{\nano\meter} and \SI{854}{\nano\meter} for repumping the metastable D-levels is used during all cooling stages, state detection, and to reset the electronic state before the next experiment, respectively. We repeat this process 50 times to collect statistics. All of these steps are standard in quantum information processing with trapped ions \cite{Haffner2008}.


\bibliography{bibliography}

\begin{thebibliography}{39}%
\makeatletter
\providecommand \@ifxundefined [1]{%
 \@ifx{#1\undefined}
}%
\providecommand \@ifnum [1]{%
 \ifnum #1\expandafter \@firstoftwo
 \else \expandafter \@secondoftwo
 \fi
}%
\providecommand \@ifx [1]{%
 \ifx #1\expandafter \@firstoftwo
 \else \expandafter \@secondoftwo
 \fi
}%
\providecommand \natexlab [1]{#1}%
\providecommand \enquote  [1]{``#1''}%
\providecommand \bibnamefont  [1]{#1}%
\providecommand \bibfnamefont [1]{#1}%
\providecommand \citenamefont [1]{#1}%
\providecommand \href@noop [0]{\@secondoftwo}%
\providecommand \href [0]{\begingroup \@sanitize@url \@href}%
\providecommand \@href[1]{\@@startlink{#1}\@@href}%
\providecommand \@@href[1]{\endgroup#1\@@endlink}%
\providecommand \@sanitize@url [0]{\catcode `\\12\catcode `\$12\catcode
  `\&12\catcode `\#12\catcode `\^12\catcode `\_12\catcode `\%12\relax}%
\providecommand \@@startlink[1]{}%
\providecommand \@@endlink[0]{}%
\providecommand \url  [0]{\begingroup\@sanitize@url \@url }%
\providecommand \@url [1]{\endgroup\@href {#1}{\urlprefix }}%
\providecommand \urlprefix  [0]{URL }%
\providecommand \Eprint [0]{\href }%
\providecommand \doibase [0]{https://doi.org/}%
\providecommand \selectlanguage [0]{\@gobble}%
\providecommand \bibinfo  [0]{\@secondoftwo}%
\providecommand \bibfield  [0]{\@secondoftwo}%
\providecommand \translation [1]{[#1]}%
\providecommand \BibitemOpen [0]{}%
\providecommand \bibitemStop [0]{}%
\providecommand \bibitemNoStop [0]{.\EOS\space}%
\providecommand \EOS [0]{\spacefactor3000\relax}%
\providecommand \BibitemShut  [1]{\csname bibitem#1\endcsname}%
\let\auto@bib@innerbib\@empty
\bibitem [{\citenamefont {Phillips}\ \emph {et~al.}(1985)\citenamefont
  {Phillips}, \citenamefont {Prodan},\ and\ \citenamefont
  {Metcalf}}]{Phillips85}%
  \BibitemOpen
  \bibfield  {author} {\bibinfo {author} {\bibfnamefont {W.~D.}\ \bibnamefont
  {Phillips}}, \bibinfo {author} {\bibfnamefont {J.~V.}\ \bibnamefont
  {Prodan}},\ and\ \bibinfo {author} {\bibfnamefont {H.~J.}\ \bibnamefont
  {Metcalf}},\ }\bibfield  {title} {\bibinfo {title} {Laser cooling and
  electromagnetic trapping of neutral atoms},\ }\href
  {https://doi.org/10.1364/JOSAB.2.001751} {\bibfield  {journal} {\bibinfo
  {journal} {J. Opt. Soc. Am. B}\ }\textbf {\bibinfo {volume} {2}},\ \bibinfo
  {pages} {1751} (\bibinfo {year} {1985})}\BibitemShut {NoStop}%
\bibitem [{\citenamefont {Grimm}\ \emph {et~al.}(2000)\citenamefont {Grimm},
  \citenamefont {Weidemüller},\ and\ \citenamefont {Ovchinnikov}}]{Grimm95}%
  \BibitemOpen
  \bibfield  {author} {\bibinfo {author} {\bibfnamefont {R.}~\bibnamefont
  {Grimm}}, \bibinfo {author} {\bibfnamefont {M.}~\bibnamefont
  {Weidemüller}},\ and\ \bibinfo {author} {\bibfnamefont {Y.~B.}\ \bibnamefont
  {Ovchinnikov}},\ }\bibfield  {title} {\bibinfo {title} {Optical dipole traps
  for neutral atoms}\ }(\bibinfo  {publisher} {Academic Press},\ \bibinfo
  {year} {2000})\ pp.\ \bibinfo {pages} {95--170}\BibitemShut {NoStop}%
\bibitem [{\citenamefont {Eschner}\ \emph {et~al.}(2003)\citenamefont
  {Eschner}, \citenamefont {Morigi}, \citenamefont {Schmidt-Kaler},\ and\
  \citenamefont {Blatt}}]{Eschner03}%
  \BibitemOpen
  \bibfield  {author} {\bibinfo {author} {\bibfnamefont {J.}~\bibnamefont
  {Eschner}}, \bibinfo {author} {\bibfnamefont {G.}~\bibnamefont {Morigi}},
  \bibinfo {author} {\bibfnamefont {F.}~\bibnamefont {Schmidt-Kaler}},\ and\
  \bibinfo {author} {\bibfnamefont {R.}~\bibnamefont {Blatt}},\ }\bibfield
  {title} {\bibinfo {title} {Laser cooling of trapped ions},\ }\href
  {https://doi.org/10.1364/JOSAB.20.001003} {\bibfield  {journal} {\bibinfo
  {journal} {J. Opt. Soc. Am. B}\ }\textbf {\bibinfo {volume} {20}},\ \bibinfo
  {pages} {1003} (\bibinfo {year} {2003})}\BibitemShut {NoStop}%
\bibitem [{\citenamefont {Blatt}\ and\ \citenamefont {Roos}(2012)}]{Blatt2012}%
  \BibitemOpen
  \bibfield  {author} {\bibinfo {author} {\bibfnamefont {R.}~\bibnamefont
  {Blatt}}\ and\ \bibinfo {author} {\bibfnamefont {C.~F.}\ \bibnamefont
  {Roos}},\ }\bibfield  {title} {\bibinfo {title} {Quantum simulations with
  trapped ions},\ }\href {https://doi.org/10.1038/nphys2252
  http://10.0.4.14/nphys2252} {\bibfield  {journal} {\bibinfo  {journal}
  {Nature Physics}\ }\textbf {\bibinfo {volume} {8}},\ \bibinfo {pages} {277}
  (\bibinfo {year} {2012})}\BibitemShut {NoStop}%
\bibitem [{\citenamefont {Ludlow}\ \emph {et~al.}(2015)\citenamefont {Ludlow},
  \citenamefont {Boyd}, \citenamefont {Ye}, \citenamefont {Peik},\ and\
  \citenamefont {Schmidt}}]{Ludlow2015}%
  \BibitemOpen
  \bibfield  {author} {\bibinfo {author} {\bibfnamefont {A.~D.}\ \bibnamefont
  {Ludlow}}, \bibinfo {author} {\bibfnamefont {M.~M.}\ \bibnamefont {Boyd}},
  \bibinfo {author} {\bibfnamefont {J.}~\bibnamefont {Ye}}, \bibinfo {author}
  {\bibfnamefont {E.}~\bibnamefont {Peik}},\ and\ \bibinfo {author}
  {\bibfnamefont {P.~O.}\ \bibnamefont {Schmidt}},\ }\bibfield  {title}
  {\bibinfo {title} {Optical atomic clocks},\ }\href
  {https://doi.org/10.1103/REVMODPHYS.87.637/FIGURES/19/MEDIUM} {\bibfield
  {journal} {\bibinfo  {journal} {Reviews of Modern Physics}\ }\textbf
  {\bibinfo {volume} {87}},\ \bibinfo {pages} {637} (\bibinfo {year}
  {2015})}\BibitemShut {NoStop}%
\bibitem [{\citenamefont {Ray}\ \emph {et~al.}(2020)\citenamefont {Ray},
  \citenamefont {Gupta}, \citenamefont {Gokhroo}, \citenamefont {Everett},
  \citenamefont {Nieddu}, \citenamefont {Rajasree}, \citenamefont {Chormaic},\
  and\ \citenamefont {Chormaic}}]{Ray2020}%
  \BibitemOpen
  \bibfield  {author} {\bibinfo {author} {\bibfnamefont {T.}~\bibnamefont
  {Ray}}, \bibinfo {author} {\bibfnamefont {R.~K.}\ \bibnamefont {Gupta}},
  \bibinfo {author} {\bibfnamefont {V.}~\bibnamefont {Gokhroo}}, \bibinfo
  {author} {\bibfnamefont {J.~L.}\ \bibnamefont {Everett}}, \bibinfo {author}
  {\bibfnamefont {T.}~\bibnamefont {Nieddu}}, \bibinfo {author} {\bibfnamefont
  {K.~S.}\ \bibnamefont {Rajasree}}, \bibinfo {author} {\bibfnamefont {S.~N.}\
  \bibnamefont {Chormaic}},\ and\ \bibinfo {author} {\bibfnamefont {S.~N.}\
  \bibnamefont {Chormaic}},\ }\bibfield  {title} {\bibinfo {title} {Observation
  of the 87rb 5s1/2 to 4d3/2 electric quadrupole transition at 516.6 nm
  mediated via an optical nanofibre},\ }\href
  {https://doi.org/10.1088/1367-2630/AB8265} {\bibfield  {journal} {\bibinfo
  {journal} {New Journal of Physics}\ }\textbf {\bibinfo {volume} {22}},\
  \bibinfo {pages} {062001} (\bibinfo {year} {2020})}\BibitemShut {NoStop}%
\bibitem [{\citenamefont {Tojo}\ \emph {et~al.}(2004)\citenamefont {Tojo},
  \citenamefont {Hasuo},\ and\ \citenamefont {Fujimoto}}]{Tojo2004}%
  \BibitemOpen
  \bibfield  {author} {\bibinfo {author} {\bibfnamefont {S.}~\bibnamefont
  {Tojo}}, \bibinfo {author} {\bibfnamefont {M.}~\bibnamefont {Hasuo}},\ and\
  \bibinfo {author} {\bibfnamefont {T.}~\bibnamefont {Fujimoto}},\ }\bibfield
  {title} {\bibinfo {title} {Absorption enhancement of an electric quadrupole
  transition of cesium atoms in an evanescent field},\ }\href
  {https://doi.org/10.1103/PHYSREVLETT.92.053001/FIGURES/4/MEDIUM} {\bibfield
  {journal} {\bibinfo  {journal} {Physical Review Letters}\ }\textbf {\bibinfo
  {volume} {92}},\ \bibinfo {pages} {4} (\bibinfo {year} {2004})}\BibitemShut
  {NoStop}%
\bibitem [{\citenamefont {Zhang}\ \emph {et~al.}(2020)\citenamefont {Zhang},
  \citenamefont {Wang}, \citenamefont {Zhu}, \citenamefont {Wang},
  \citenamefont {Zhang}, \citenamefont {Gao},\ and\ \citenamefont
  {Zhang}}]{Zhang2020}%
  \BibitemOpen
  \bibfield  {author} {\bibinfo {author} {\bibfnamefont {Q.}~\bibnamefont
  {Zhang}}, \bibinfo {author} {\bibfnamefont {Y.}~\bibnamefont {Wang}},
  \bibinfo {author} {\bibfnamefont {C.}~\bibnamefont {Zhu}}, \bibinfo {author}
  {\bibfnamefont {Y.}~\bibnamefont {Wang}}, \bibinfo {author} {\bibfnamefont
  {X.}~\bibnamefont {Zhang}}, \bibinfo {author} {\bibfnamefont
  {K.}~\bibnamefont {Gao}},\ and\ \bibinfo {author} {\bibfnamefont
  {W.}~\bibnamefont {Zhang}},\ }\bibfield  {title} {\bibinfo {title} {Precision
  measurements with cold atoms and trapped ions},\ }\href
  {https://doi.org/10.1088/1674-1056/aba9c6} {\bibfield  {journal} {\bibinfo
  {journal} {Chinese Physics B}\ }\textbf {\bibinfo {volume} {29}},\ \bibinfo
  {pages} {093203} (\bibinfo {year} {2020})}\BibitemShut {NoStop}%
\bibitem [{\citenamefont {Leibfried}\ \emph {et~al.}(2003)\citenamefont
  {Leibfried}, \citenamefont {Blatt}, \citenamefont {Monroe},\ and\
  \citenamefont {Wineland}}]{Leibfried2003}%
  \BibitemOpen
  \bibfield  {author} {\bibinfo {author} {\bibfnamefont {D.}~\bibnamefont
  {Leibfried}}, \bibinfo {author} {\bibfnamefont {R.}~\bibnamefont {Blatt}},
  \bibinfo {author} {\bibfnamefont {C.}~\bibnamefont {Monroe}},\ and\ \bibinfo
  {author} {\bibfnamefont {D.}~\bibnamefont {Wineland}},\ }\bibfield  {title}
  {\bibinfo {title} {{Quantum dynamics of single trapped ions}},\ }\href
  {https://doi.org/10.1103/RevModPhys.75.281} {\bibfield  {journal} {\bibinfo
  {journal} {Reviews of Modern Physics}\ }\textbf {\bibinfo {volume} {75}},\
  \bibinfo {pages} {281} (\bibinfo {year} {2003})}\BibitemShut {NoStop}%
\bibitem [{\citenamefont {Bruzewicz}\ \emph {et~al.}(2019)\citenamefont
  {Bruzewicz}, \citenamefont {Chiaverini}, \citenamefont {McConnell},\ and\
  \citenamefont {Sage}}]{Bruzewicz2019}%
  \BibitemOpen
  \bibfield  {author} {\bibinfo {author} {\bibfnamefont {C.~D.}\ \bibnamefont
  {Bruzewicz}}, \bibinfo {author} {\bibfnamefont {J.}~\bibnamefont
  {Chiaverini}}, \bibinfo {author} {\bibfnamefont {R.}~\bibnamefont
  {McConnell}},\ and\ \bibinfo {author} {\bibfnamefont {J.~M.}\ \bibnamefont
  {Sage}},\ }\bibfield  {title} {\bibinfo {title} {{Trapped-ion quantum
  computing: Progress and challenges}},\ }\href
  {https://doi.org/10.1063/1.5088164} {\bibfield  {journal} {\bibinfo
  {journal} {Applied Physics Reviews}\ }\textbf {\bibinfo {volume} {6}},\
  \bibinfo {pages} {21314} (\bibinfo {year} {2019})},\ \Eprint
  {https://arxiv.org/abs/1904.04178} {arXiv:1904.04178} \BibitemShut {NoStop}%
\bibitem [{\citenamefont {Mundt}\ \emph {et~al.}(2002)\citenamefont {Mundt},
  \citenamefont {Kreuter}, \citenamefont {Becher}, \citenamefont {Leibfried},
  \citenamefont {Eschner}, \citenamefont {Schmidt-Kaler},\ and\ \citenamefont
  {Blatt}}]{Mundt2002}%
  \BibitemOpen
  \bibfield  {author} {\bibinfo {author} {\bibfnamefont {A.~B.}\ \bibnamefont
  {Mundt}}, \bibinfo {author} {\bibfnamefont {A.}~\bibnamefont {Kreuter}},
  \bibinfo {author} {\bibfnamefont {C.}~\bibnamefont {Becher}}, \bibinfo
  {author} {\bibfnamefont {D.}~\bibnamefont {Leibfried}}, \bibinfo {author}
  {\bibfnamefont {J.}~\bibnamefont {Eschner}}, \bibinfo {author} {\bibfnamefont
  {F.}~\bibnamefont {Schmidt-Kaler}},\ and\ \bibinfo {author} {\bibfnamefont
  {R.}~\bibnamefont {Blatt}},\ }\bibfield  {title} {\bibinfo {title} {{Coupling
  a Single Atomic Quantum Bit to a High Finesse Optical Cavity}},\ }\href
  {https://doi.org/10.1103/PhysRevLett.89.103001} {\bibfield  {journal}
  {\bibinfo  {journal} {Physical Review Letters}\ }\textbf {\bibinfo {volume}
  {89}},\ \bibinfo {pages} {103001} (\bibinfo {year} {2002})},\ \Eprint
  {https://arxiv.org/abs/0202112} {arXiv:0202112 [quant-ph]} \BibitemShut
  {NoStop}%
\bibitem [{\citenamefont {Schmiegelow}\ \emph
  {et~al.}(2016{\natexlab{a}})\citenamefont {Schmiegelow}, \citenamefont
  {Schulz}, \citenamefont {Kaufmann}, \citenamefont {Ruster}, \citenamefont
  {Poschinger},\ and\ \citenamefont {Schmidt-Kaler}}]{Schmiegelow2016a}%
  \BibitemOpen
  \bibfield  {author} {\bibinfo {author} {\bibfnamefont {C.~T.}\ \bibnamefont
  {Schmiegelow}}, \bibinfo {author} {\bibfnamefont {J.}~\bibnamefont {Schulz}},
  \bibinfo {author} {\bibfnamefont {H.}~\bibnamefont {Kaufmann}}, \bibinfo
  {author} {\bibfnamefont {T.}~\bibnamefont {Ruster}}, \bibinfo {author}
  {\bibfnamefont {U.~G.}\ \bibnamefont {Poschinger}},\ and\ \bibinfo {author}
  {\bibfnamefont {F.}~\bibnamefont {Schmidt-Kaler}},\ }\bibfield  {title}
  {\bibinfo {title} {{Transfer of optical orbital angular momentum to a bound
  electron}},\ }\href {https://doi.org/10.1038/ncomms12998} {\bibfield
  {journal} {\bibinfo  {journal} {Nature Communications}\ }\textbf {\bibinfo
  {volume} {7}},\ \bibinfo {pages} {1} (\bibinfo {year}
  {2016}{\natexlab{a}})}\BibitemShut {NoStop}%
\bibitem [{\citenamefont {Drechsler}\ \emph {et~al.}(2021)\citenamefont
  {Drechsler}, \citenamefont {Wolf}, \citenamefont {Schmiegelow},\ and\
  \citenamefont {Schmidt-Kaler}}]{Drechsler2021}%
  \BibitemOpen
  \bibfield  {author} {\bibinfo {author} {\bibfnamefont {M.}~\bibnamefont
  {Drechsler}}, \bibinfo {author} {\bibfnamefont {S.}~\bibnamefont {Wolf}},
  \bibinfo {author} {\bibfnamefont {C.~T.}\ \bibnamefont {Schmiegelow}},\ and\
  \bibinfo {author} {\bibfnamefont {F.}~\bibnamefont {Schmidt-Kaler}},\
  }\bibfield  {title} {\bibinfo {title} {{Optical Superresolution Sensing of a
  Trapped Ion's Wave Packet Size}},\ }\href
  {https://doi.org/10.1103/PhysRevLett.127.143602} {\bibfield  {journal}
  {\bibinfo  {journal} {Physical Review Letters}\ }\textbf {\bibinfo {volume}
  {127}},\ \bibinfo {pages} {143602} (\bibinfo {year} {2021})},\ \Eprint
  {https://arxiv.org/abs/2104.07095} {arXiv:2104.07095} \BibitemShut {NoStop}%
\bibitem [{\citenamefont {Gross}\ and\ \citenamefont
  {Bloch}(2017)}]{Gross2017}%
  \BibitemOpen
  \bibfield  {author} {\bibinfo {author} {\bibfnamefont {C.}~\bibnamefont
  {Gross}}\ and\ \bibinfo {author} {\bibfnamefont {I.}~\bibnamefont {Bloch}},\
  }\bibfield  {title} {\bibinfo {title} {Quantum simulations with ultracold
  atoms in optical lattices},\ }\href {https://doi.org/10.1126/science.aal3837}
  {\bibfield  {journal} {\bibinfo  {journal} {Science}\ }\textbf {\bibinfo
  {volume} {357}},\ \bibinfo {pages} {995} (\bibinfo {year} {2017})},\ \Eprint
  {https://arxiv.org/abs/https://www.science.org/doi/pdf/10.1126/science.aal3837}
  {https://www.science.org/doi/pdf/10.1126/science.aal3837} \BibitemShut
  {NoStop}%
\bibitem [{\citenamefont {Subhankar}\ \emph {et~al.}(2019)\citenamefont
  {Subhankar}, \citenamefont {Wang}, \citenamefont {Tsui}, \citenamefont
  {Rolston},\ and\ \citenamefont {Porto}}]{Subhankar2019}%
  \BibitemOpen
  \bibfield  {author} {\bibinfo {author} {\bibfnamefont {S.}~\bibnamefont
  {Subhankar}}, \bibinfo {author} {\bibfnamefont {Y.}~\bibnamefont {Wang}},
  \bibinfo {author} {\bibfnamefont {T.-C.}\ \bibnamefont {Tsui}}, \bibinfo
  {author} {\bibfnamefont {S.~L.}\ \bibnamefont {Rolston}},\ and\ \bibinfo
  {author} {\bibfnamefont {J.~V.}\ \bibnamefont {Porto}},\ }\bibfield  {title}
  {\bibinfo {title} {Nanoscale atomic density microscopy},\ }\href
  {https://doi.org/10.1103/PhysRevX.9.021002} {\bibfield  {journal} {\bibinfo
  {journal} {Phys. Rev. X}\ }\textbf {\bibinfo {volume} {9}},\ \bibinfo {pages}
  {021002} (\bibinfo {year} {2019})}\BibitemShut {NoStop}%
\bibitem [{\citenamefont {Mehta}\ \emph {et~al.}(2019)\citenamefont {Mehta},
  \citenamefont {Zhang}, \citenamefont {Miller},\ and\ \citenamefont
  {Home}}]{Mehta2019}%
  \BibitemOpen
  \bibfield  {author} {\bibinfo {author} {\bibfnamefont {K.~K.}\ \bibnamefont
  {Mehta}}, \bibinfo {author} {\bibfnamefont {C.}~\bibnamefont {Zhang}},
  \bibinfo {author} {\bibfnamefont {S.}~\bibnamefont {Miller}},\ and\ \bibinfo
  {author} {\bibfnamefont {J.~P.}\ \bibnamefont {Home}},\ }\bibfield  {title}
  {\bibinfo {title} {{Towards fast and scalable trapped-ion quantum logic with
  integrated photonics}},\ }in\ \href {https://doi.org/10.1117/12.2507647}
  {\emph {\bibinfo {booktitle} {Advances in Photonics of Quantum Computing,
  Memory, and Communication XII}}},\ Vol.\ \bibinfo {volume} {10933},\ \bibinfo
  {editor} {edited by\ \bibinfo {editor} {\bibfnamefont {P.~R.}\ \bibnamefont
  {Hemmer}}, \bibinfo {editor} {\bibfnamefont {A.~L.}\ \bibnamefont
  {Migdall}},\ and\ \bibinfo {editor} {\bibfnamefont {Z.~U.}\ \bibnamefont
  {Hasan}}},\ \bibinfo {organization} {International Society for Optics and
  Photonics}\ (\bibinfo  {publisher} {SPIE},\ \bibinfo {year} {2019})\ pp.\
  \bibinfo {pages} {24--34}\BibitemShut {NoStop}%
\bibitem [{\citenamefont {S{\o}rensen}\ and\ \citenamefont
  {M{\o}lmer}(2000)}]{Sorensen2000}%
  \BibitemOpen
  \bibfield  {author} {\bibinfo {author} {\bibfnamefont {A.}~\bibnamefont
  {S{\o}rensen}}\ and\ \bibinfo {author} {\bibfnamefont {K.}~\bibnamefont
  {M{\o}lmer}},\ }\bibfield  {title} {\bibinfo {title} {{Entanglement and
  quantum computation with ions in thermal motion}},\ }\href
  {https://doi.org/10.1103/PhysRevA.62.022311} {\bibfield  {journal} {\bibinfo
  {journal} {Physical Review A - Atomic, Molecular, and Optical Physics}\
  }\textbf {\bibinfo {volume} {62}},\ \bibinfo {pages} {11} (\bibinfo {year}
  {2000})},\ \Eprint {https://arxiv.org/abs/0002024} {arXiv:0002024 [quant-ph]}
  \BibitemShut {NoStop}%
\bibitem [{\citenamefont {de~Neeve}\ \emph {et~al.}(2022)\citenamefont
  {de~Neeve}, \citenamefont {Nguyen}, \citenamefont {Behrle},\ and\
  \citenamefont {Home}}]{deNeeves2022}%
  \BibitemOpen
  \bibfield  {author} {\bibinfo {author} {\bibfnamefont {B.}~\bibnamefont
  {de~Neeve}}, \bibinfo {author} {\bibfnamefont {T.~L.}\ \bibnamefont
  {Nguyen}}, \bibinfo {author} {\bibfnamefont {T.}~\bibnamefont {Behrle}},\
  and\ \bibinfo {author} {\bibfnamefont {J.~P.}\ \bibnamefont {Home}},\
  }\bibfield  {title} {\bibinfo {title} {Error correction of a logical grid
  state qubit by dissipative pumping},\ }\href
  {https://doi.org/10.1038/s41567-021-01487-7} {\bibfield  {journal} {\bibinfo
  {journal} {Nature Physics 2022 18:3}\ }\textbf {\bibinfo {volume} {18}},\
  \bibinfo {pages} {296} (\bibinfo {year} {2022})}\BibitemShut {NoStop}%
\bibitem [{\citenamefont {Huntemann}\ \emph {et~al.}(2012)\citenamefont
  {Huntemann}, \citenamefont {Okhapkin}, \citenamefont {Lipphardt},
  \citenamefont {Weyers}, \citenamefont {Tamm},\ and\ \citenamefont
  {Peik}}]{Huntemann2012}%
  \BibitemOpen
  \bibfield  {author} {\bibinfo {author} {\bibfnamefont {N.}~\bibnamefont
  {Huntemann}}, \bibinfo {author} {\bibfnamefont {M.}~\bibnamefont {Okhapkin}},
  \bibinfo {author} {\bibfnamefont {B.}~\bibnamefont {Lipphardt}}, \bibinfo
  {author} {\bibfnamefont {S.}~\bibnamefont {Weyers}}, \bibinfo {author}
  {\bibfnamefont {C.}~\bibnamefont {Tamm}},\ and\ \bibinfo {author}
  {\bibfnamefont {E.}~\bibnamefont {Peik}},\ }\bibfield  {title} {\bibinfo
  {title} {{High-accuracy optical clock based on the octupole transition in
  Yb+171}},\ }\href
  {https://doi.org/10.1103/PHYSREVLETT.108.090801/FIGURES/4/MEDIUM} {\bibfield
  {journal} {\bibinfo  {journal} {Physical Review Letters}\ }\textbf {\bibinfo
  {volume} {108}},\ \bibinfo {pages} {090801} (\bibinfo {year} {2012})},\
  \Eprint {https://arxiv.org/abs/1111.2446} {arXiv:1111.2446} \BibitemShut
  {NoStop}%
\bibitem [{\citenamefont {Schmiegelow}\ \emph
  {et~al.}(2016{\natexlab{b}})\citenamefont {Schmiegelow}, \citenamefont
  {Kaufmann}, \citenamefont {Ruster}, \citenamefont {Schulz}, \citenamefont
  {Kaushal}, \citenamefont {Hettrich}, \citenamefont {Schmidt-Kaler},\ and\
  \citenamefont {Poschinger}}]{Schmiegelow2016}%
  \BibitemOpen
  \bibfield  {author} {\bibinfo {author} {\bibfnamefont {C.~T.}\ \bibnamefont
  {Schmiegelow}}, \bibinfo {author} {\bibfnamefont {H.}~\bibnamefont
  {Kaufmann}}, \bibinfo {author} {\bibfnamefont {T.}~\bibnamefont {Ruster}},
  \bibinfo {author} {\bibfnamefont {J.}~\bibnamefont {Schulz}}, \bibinfo
  {author} {\bibfnamefont {V.}~\bibnamefont {Kaushal}}, \bibinfo {author}
  {\bibfnamefont {M.}~\bibnamefont {Hettrich}}, \bibinfo {author}
  {\bibfnamefont {F.}~\bibnamefont {Schmidt-Kaler}},\ and\ \bibinfo {author}
  {\bibfnamefont {U.~G.}\ \bibnamefont {Poschinger}},\ }\bibfield  {title}
  {\bibinfo {title} {{Phase-Stable Free-Space Optical Lattices for Trapped
  Ions}},\ }\href {https://doi.org/10.1103/PhysRevLett.116.033002} {\bibfield
  {journal} {\bibinfo  {journal} {Physical Review Letters}\ }\textbf {\bibinfo
  {volume} {116}},\ \bibinfo {pages} {033002} (\bibinfo {year}
  {2016}{\natexlab{b}})},\ \Eprint {https://arxiv.org/abs/1507.05207}
  {arXiv:1507.05207} \BibitemShut {NoStop}%
\bibitem [{\citenamefont {Delaubenfels}\ \emph {et~al.}(2015)\citenamefont
  {Delaubenfels}, \citenamefont {Burkhardt}, \citenamefont {Vittorini},
  \citenamefont {Merrill}, \citenamefont {Brown},\ and\ \citenamefont
  {Amini}}]{Delaubenfels2015}%
  \BibitemOpen
  \bibfield  {author} {\bibinfo {author} {\bibfnamefont {T.~E.}\ \bibnamefont
  {Delaubenfels}}, \bibinfo {author} {\bibfnamefont {K.~A.}\ \bibnamefont
  {Burkhardt}}, \bibinfo {author} {\bibfnamefont {G.}~\bibnamefont
  {Vittorini}}, \bibinfo {author} {\bibfnamefont {J.~T.}\ \bibnamefont
  {Merrill}}, \bibinfo {author} {\bibfnamefont {K.~R.}\ \bibnamefont {Brown}},\
  and\ \bibinfo {author} {\bibfnamefont {J.~M.}\ \bibnamefont {Amini}},\
  }\bibfield  {title} {\bibinfo {title} {{Modulating carrier and sideband
  coupling strengths in a standing-wave gate beam}},\ }\href
  {https://doi.org/10.1103/PhysRevA.92.061402} {\bibfield  {journal} {\bibinfo
  {journal} {Physical Review A - Atomic, Molecular, and Optical Physics}\
  }\textbf {\bibinfo {volume} {92}},\ \bibinfo {pages} {061402} (\bibinfo
  {year} {2015})}\BibitemShut {NoStop}%
\bibitem [{\citenamefont {Mehta}\ \emph {et~al.}(2016)\citenamefont {Mehta},
  \citenamefont {Bruzewicz}, \citenamefont {McConnell}, \citenamefont {Ram},
  \citenamefont {Sage},\ and\ \citenamefont {Chiaverini}}]{Mehta2016}%
  \BibitemOpen
  \bibfield  {author} {\bibinfo {author} {\bibfnamefont {K.~K.}\ \bibnamefont
  {Mehta}}, \bibinfo {author} {\bibfnamefont {C.~D.}\ \bibnamefont
  {Bruzewicz}}, \bibinfo {author} {\bibfnamefont {R.}~\bibnamefont
  {McConnell}}, \bibinfo {author} {\bibfnamefont {R.~J.}\ \bibnamefont {Ram}},
  \bibinfo {author} {\bibfnamefont {J.~M.}\ \bibnamefont {Sage}},\ and\
  \bibinfo {author} {\bibfnamefont {J.}~\bibnamefont {Chiaverini}},\ }\bibfield
   {title} {\bibinfo {title} {{Integrated optical addressing of an ion
  qubit}},\ }\href {https://doi.org/10.1038/nnano.2016.139} {\bibfield
  {journal} {\bibinfo  {journal} {Nature Nanotechnology}\ }\textbf {\bibinfo
  {volume} {11}},\ \bibinfo {pages} {1066} (\bibinfo {year}
  {2016})}\BibitemShut {NoStop}%
\bibitem [{\citenamefont {Niffenegger}\ \emph {et~al.}(2020)\citenamefont
  {Niffenegger}, \citenamefont {Stuart}, \citenamefont {Sorace-Agaskar},
  \citenamefont {Kharas}, \citenamefont {Bramhavar}, \citenamefont {Bruzewicz},
  \citenamefont {Loh}, \citenamefont {Maxson}, \citenamefont {McConnell},
  \citenamefont {Reens}, \citenamefont {West}, \citenamefont {Sage},\ and\
  \citenamefont {Chiaverini}}]{Niffenegger2020}%
  \BibitemOpen
  \bibfield  {author} {\bibinfo {author} {\bibfnamefont {R.~J.}\ \bibnamefont
  {Niffenegger}}, \bibinfo {author} {\bibfnamefont {J.}~\bibnamefont {Stuart}},
  \bibinfo {author} {\bibfnamefont {C.}~\bibnamefont {Sorace-Agaskar}},
  \bibinfo {author} {\bibfnamefont {D.}~\bibnamefont {Kharas}}, \bibinfo
  {author} {\bibfnamefont {S.}~\bibnamefont {Bramhavar}}, \bibinfo {author}
  {\bibfnamefont {C.~D.}\ \bibnamefont {Bruzewicz}}, \bibinfo {author}
  {\bibfnamefont {W.}~\bibnamefont {Loh}}, \bibinfo {author} {\bibfnamefont
  {R.~T.}\ \bibnamefont {Maxson}}, \bibinfo {author} {\bibfnamefont
  {R.}~\bibnamefont {McConnell}}, \bibinfo {author} {\bibfnamefont
  {D.}~\bibnamefont {Reens}}, \bibinfo {author} {\bibfnamefont {G.~N.}\
  \bibnamefont {West}}, \bibinfo {author} {\bibfnamefont {J.~M.}\ \bibnamefont
  {Sage}},\ and\ \bibinfo {author} {\bibfnamefont {J.}~\bibnamefont
  {Chiaverini}},\ }\bibfield  {title} {\bibinfo {title} {{Integrated
  multi-wavelength control of an ion qubit}},\ }\href
  {https://doi.org/10.1038/s41586-020-2811-x} {\bibfield  {journal} {\bibinfo
  {journal} {Nature}\ }\textbf {\bibinfo {volume} {586}},\ \bibinfo {pages}
  {538} (\bibinfo {year} {2020})}\BibitemShut {NoStop}%
\bibitem [{\citenamefont {Mehta}\ \emph {et~al.}(2020)\citenamefont {Mehta},
  \citenamefont {Zhang}, \citenamefont {Malinowski}, \citenamefont {Nguyen},
  \citenamefont {Stadler},\ and\ \citenamefont {Home}}]{Mehta2020}%
  \BibitemOpen
  \bibfield  {author} {\bibinfo {author} {\bibfnamefont {K.~K.}\ \bibnamefont
  {Mehta}}, \bibinfo {author} {\bibfnamefont {C.}~\bibnamefont {Zhang}},
  \bibinfo {author} {\bibfnamefont {M.}~\bibnamefont {Malinowski}}, \bibinfo
  {author} {\bibfnamefont {T.-L.}\ \bibnamefont {Nguyen}}, \bibinfo {author}
  {\bibfnamefont {M.}~\bibnamefont {Stadler}},\ and\ \bibinfo {author}
  {\bibfnamefont {J.~P.}\ \bibnamefont {Home}},\ }\bibfield  {title} {\bibinfo
  {title} {{Integrated optical multi-ion quantum logic}},\ }\href
  {https://doi.org/10.1038/s41586-020-2823-6} {\bibfield  {journal} {\bibinfo
  {journal} {Nature}\ }\textbf {\bibinfo {volume} {586}},\ \bibinfo {pages}
  {533} (\bibinfo {year} {2020})}\BibitemShut {NoStop}%
\bibitem [{\citenamefont {Ivory}\ \emph {et~al.}(2020)\citenamefont {Ivory},
  \citenamefont {Setzer}, \citenamefont {Karl}, \citenamefont {McGuinness},
  \citenamefont {DeRose}, \citenamefont {Blain}, \citenamefont {Stick},
  \citenamefont {Gehl},\ and\ \citenamefont {Parazzoli}}]{Ivory2020}%
  \BibitemOpen
  \bibfield  {author} {\bibinfo {author} {\bibfnamefont {M.}~\bibnamefont
  {Ivory}}, \bibinfo {author} {\bibfnamefont {W.~J.}\ \bibnamefont {Setzer}},
  \bibinfo {author} {\bibfnamefont {N.}~\bibnamefont {Karl}}, \bibinfo {author}
  {\bibfnamefont {H.}~\bibnamefont {McGuinness}}, \bibinfo {author}
  {\bibfnamefont {C.}~\bibnamefont {DeRose}}, \bibinfo {author} {\bibfnamefont
  {M.}~\bibnamefont {Blain}}, \bibinfo {author} {\bibfnamefont
  {D.}~\bibnamefont {Stick}}, \bibinfo {author} {\bibfnamefont
  {M.}~\bibnamefont {Gehl}},\ and\ \bibinfo {author} {\bibfnamefont {L.~P.}\
  \bibnamefont {Parazzoli}},\ }\bibfield  {title} {\bibinfo {title}
  {{Integrated optical addressing of a trapped ytterbium ion}},\ }\href
  {https://doi.org/10.1103/PHYSREVX.11.041033/FIGURES/8/MEDIUM} {\bibfield
  {journal} {\bibinfo  {journal} {Physical Review X}\ }\textbf {\bibinfo
  {volume} {11}},\ \bibinfo {pages} {041033} (\bibinfo {year} {2020})},\
  \Eprint {https://arxiv.org/abs/2011.12376} {arXiv:2011.12376} \BibitemShut
  {NoStop}%
\bibitem [{\citenamefont {Zhang}(2022)}]{Zhang2022}%
  \BibitemOpen
  \bibfield  {author} {\bibinfo {author} {\bibfnamefont {C.}~\bibnamefont
  {Zhang}},\ }\emph {\bibinfo {title} {Scalable technologies for
  surface-electrode ion traps}},\ \href
  {https://doi.org/10.3929/ETHZ-B-000536415} {Ph.D. thesis},\ \bibinfo
  {school} {ETH Zurich} (\bibinfo {year} {2022})\BibitemShut {NoStop}%
\bibitem [{\citenamefont {H{\"{a}}ffner}\ \emph {et~al.}(2003)\citenamefont
  {H{\"{a}}ffner}, \citenamefont {Gulde}, \citenamefont {Riebe}, \citenamefont
  {Lancaster}, \citenamefont {Becher}, \citenamefont {Eschner}, \citenamefont
  {Schmidt-Kaler},\ and\ \citenamefont {Blatt}}]{Haffner2003}%
  \BibitemOpen
  \bibfield  {author} {\bibinfo {author} {\bibfnamefont {H.}~\bibnamefont
  {H{\"{a}}ffner}}, \bibinfo {author} {\bibfnamefont {S.}~\bibnamefont
  {Gulde}}, \bibinfo {author} {\bibfnamefont {M.}~\bibnamefont {Riebe}},
  \bibinfo {author} {\bibfnamefont {G.}~\bibnamefont {Lancaster}}, \bibinfo
  {author} {\bibfnamefont {C.}~\bibnamefont {Becher}}, \bibinfo {author}
  {\bibfnamefont {J.}~\bibnamefont {Eschner}}, \bibinfo {author} {\bibfnamefont
  {F.}~\bibnamefont {Schmidt-Kaler}},\ and\ \bibinfo {author} {\bibfnamefont
  {R.}~\bibnamefont {Blatt}},\ }\bibfield  {title} {\bibinfo {title}
  {{Precision Measurement and Compensation of Optical Stark Shifts for an
  Ion-Trap Quantum Processor}},\ }\href
  {https://doi.org/10.1103/PhysRevLett.90.143602} {\bibfield  {journal}
  {\bibinfo  {journal} {Physical Review Letters}\ }\textbf {\bibinfo {volume}
  {90}},\ \bibinfo {pages} {143602} (\bibinfo {year} {2003})}\BibitemShut
  {NoStop}%
\bibitem [{\citenamefont {Harlander}\ \emph {et~al.}(2010)\citenamefont
  {Harlander}, \citenamefont {Brownnutt}, \citenamefont {Hänsel},\ and\
  \citenamefont {Blatt}}]{Harlander2010}%
  \BibitemOpen
  \bibfield  {author} {\bibinfo {author} {\bibfnamefont {M.}~\bibnamefont
  {Harlander}}, \bibinfo {author} {\bibfnamefont {M.}~\bibnamefont
  {Brownnutt}}, \bibinfo {author} {\bibfnamefont {W.}~\bibnamefont {Hänsel}},\
  and\ \bibinfo {author} {\bibfnamefont {R.}~\bibnamefont {Blatt}},\ }\bibfield
   {title} {\bibinfo {title} {Trapped-ion probing of light-induced charging
  effects on dielectrics},\ }\href
  {https://doi.org/10.1088/1367-2630/12/9/093035} {\bibfield  {journal}
  {\bibinfo  {journal} {New Journal of Physics}\ }\textbf {\bibinfo {volume}
  {12}},\ \bibinfo {pages} {093035} (\bibinfo {year} {2010})}\BibitemShut
  {NoStop}%
\bibitem [{\citenamefont {Malinowski}(2021)}]{Malinowski2021}%
  \BibitemOpen
  \bibfield  {author} {\bibinfo {author} {\bibfnamefont {M.}~\bibnamefont
  {Malinowski}},\ }\emph {\bibinfo {title} {Unitary and Dissipative Trapped-Ion
  Entanglement Using Integrated Optics}},\ \href
  {https://doi.org/10.3929/ETHZ-B-000516613} {Ph.D. thesis},\ \bibinfo
  {school} {ETH Zurich} (\bibinfo {year} {2021})\BibitemShut {NoStop}%
\bibitem [{\citenamefont {Jackson}(1962)}]{Jackson1962}%
  \BibitemOpen
  \bibfield  {author} {\bibinfo {author} {\bibfnamefont {J.~D.}\ \bibnamefont
  {Jackson}},\ }\href@noop {} {\emph {\bibinfo {title} {Classical
  Electrodynamics}}}\ (\bibinfo  {publisher} {John Wiley \& Sons, Inc.},\
  \bibinfo {year} {1962})\BibitemShut {NoStop}%
\bibitem [{\citenamefont {James}(1998)}]{James1998}%
  \BibitemOpen
  \bibfield  {author} {\bibinfo {author} {\bibfnamefont {D.~F.}\ \bibnamefont
  {James}},\ }\bibfield  {title} {\bibinfo {title} {{Quantum dynamics of cold
  trapped ions with application to quantum computation}},\ }\href
  {https://doi.org/10.1007/s003400050373} {\bibfield  {journal} {\bibinfo
  {journal} {Applied Physics B: Lasers and Optics}\ }\textbf {\bibinfo {volume}
  {66}},\ \bibinfo {pages} {181} (\bibinfo {year} {1998})},\ \Eprint
  {https://arxiv.org/abs/9702053} {arXiv:9702053 [quant-ph]} \BibitemShut
  {NoStop}%
\bibitem [{\citenamefont {Barton}\ \emph {et~al.}(2000)\citenamefont {Barton},
  \citenamefont {Donald}, \citenamefont {Lucas}, \citenamefont {Stevens},
  \citenamefont {Steane},\ and\ \citenamefont {Stacey}}]{Barton2000}%
  \BibitemOpen
  \bibfield  {author} {\bibinfo {author} {\bibfnamefont {P.~A.}\ \bibnamefont
  {Barton}}, \bibinfo {author} {\bibfnamefont {C.~J.}\ \bibnamefont {Donald}},
  \bibinfo {author} {\bibfnamefont {D.~M.}\ \bibnamefont {Lucas}}, \bibinfo
  {author} {\bibfnamefont {D.~A.}\ \bibnamefont {Stevens}}, \bibinfo {author}
  {\bibfnamefont {A.~M.}\ \bibnamefont {Steane}},\ and\ \bibinfo {author}
  {\bibfnamefont {D.~N.}\ \bibnamefont {Stacey}},\ }\bibfield  {title}
  {\bibinfo {title} {Measurement of the lifetime of the {3d${^2}D_{5/2}$} state
  in {$^{40}$Ca$^+$}},\ }\href {https://doi.org/10.1103/PhysRevA.62.032503}
  {\bibfield  {journal} {\bibinfo  {journal} {Physical Review A}\ }\textbf
  {\bibinfo {volume} {62}},\ \bibinfo {pages} {032503} (\bibinfo {year}
  {2000})}\BibitemShut {NoStop}%
\bibitem [{\citenamefont {Delone}\ and\ \citenamefont
  {Krainov}(1999)}]{Delone1999}%
  \BibitemOpen
  \bibfield  {author} {\bibinfo {author} {\bibfnamefont {N.~B.}\ \bibnamefont
  {Delone}}\ and\ \bibinfo {author} {\bibfnamefont {V.~P.}\ \bibnamefont
  {Krainov}},\ }\bibfield  {title} {\bibinfo {title} {Ac stark shift of atomic
  energy levels},\ }\href {https://doi.org/10.1070/PU1999v042n07ABEH000557}
  {\bibfield  {journal} {\bibinfo  {journal} {Physics-Uspekhi}\ }\textbf
  {\bibinfo {volume} {42}},\ \bibinfo {pages} {669} (\bibinfo {year}
  {1999})}\BibitemShut {NoStop}%
\bibitem [{\citenamefont {Grau}()}]{Grau}%
  \BibitemOpen
  \bibfield  {author} {\bibinfo {author} {\bibfnamefont {M.}~\bibnamefont
  {Grau}},\ }\href {https://atomphys.org} {\bibinfo {title}
  {https://atomphys.org}}\BibitemShut {NoStop}%
\bibitem [{\citenamefont {Zhang}\ \emph {et~al.}(2013)\citenamefont {Zhang},
  \citenamefont {Yang}, \citenamefont {Lim}, \citenamefont {Lo}, \citenamefont
  {Galland}, \citenamefont {Baehr-Jones},\ and\ \citenamefont
  {Hochberg}}]{zhang2013compact}%
  \BibitemOpen
  \bibfield  {author} {\bibinfo {author} {\bibfnamefont {Y.}~\bibnamefont
  {Zhang}}, \bibinfo {author} {\bibfnamefont {S.}~\bibnamefont {Yang}},
  \bibinfo {author} {\bibfnamefont {A.~E.-J.}\ \bibnamefont {Lim}}, \bibinfo
  {author} {\bibfnamefont {G.-Q.}\ \bibnamefont {Lo}}, \bibinfo {author}
  {\bibfnamefont {C.}~\bibnamefont {Galland}}, \bibinfo {author} {\bibfnamefont
  {T.}~\bibnamefont {Baehr-Jones}},\ and\ \bibinfo {author} {\bibfnamefont
  {M.}~\bibnamefont {Hochberg}},\ }\bibfield  {title} {\bibinfo {title} {A
  compact and low loss y-junction for submicron silicon waveguide},\ }\href
  {https://doi.org/10.1364/OE.21.001310} {\bibfield  {journal} {\bibinfo
  {journal} {Opt. Express}\ }\textbf {\bibinfo {volume} {21}},\ \bibinfo
  {pages} {1310} (\bibinfo {year} {2013})}\BibitemShut {NoStop}%
\bibitem [{\citenamefont {Soldano}\ and\ \citenamefont
  {Pennings}(1995)}]{soldano1995optical}%
  \BibitemOpen
  \bibfield  {author} {\bibinfo {author} {\bibfnamefont {L.~B.}\ \bibnamefont
  {Soldano}}\ and\ \bibinfo {author} {\bibfnamefont {E.~C.}\ \bibnamefont
  {Pennings}},\ }\bibfield  {title} {\bibinfo {title} {Optical multi-mode
  interference devices based on self-imaging: principles and applications},\
  }\href@noop {} {\bibfield  {journal} {\bibinfo  {journal} {Journal of
  lightwave technology}\ }\textbf {\bibinfo {volume} {13}},\ \bibinfo {pages}
  {615} (\bibinfo {year} {1995})}\BibitemShut {NoStop}%
\bibitem [{\citenamefont {Meekhof}\ \emph {et~al.}(1996)\citenamefont
  {Meekhof}, \citenamefont {Monroe}, \citenamefont {King}, \citenamefont
  {Itano},\ and\ \citenamefont {Wineland}}]{Meekhof1996}%
  \BibitemOpen
  \bibfield  {author} {\bibinfo {author} {\bibfnamefont {D.~M.}\ \bibnamefont
  {Meekhof}}, \bibinfo {author} {\bibfnamefont {C.}~\bibnamefont {Monroe}},
  \bibinfo {author} {\bibfnamefont {B.~E.}\ \bibnamefont {King}}, \bibinfo
  {author} {\bibfnamefont {W.~M.}\ \bibnamefont {Itano}},\ and\ \bibinfo
  {author} {\bibfnamefont {D.~J.}\ \bibnamefont {Wineland}},\ }\bibfield
  {title} {\bibinfo {title} {Generation of nonclassical motional states of a
  trapped atom},\ }\href {https://doi.org/10.1103/PhysRevLett.76.1796}
  {\bibfield  {journal} {\bibinfo  {journal} {Physical Review Letters}\
  }\textbf {\bibinfo {volume} {76}},\ \bibinfo {pages} {1796} (\bibinfo {year}
  {1996})}\BibitemShut {NoStop}%
\bibitem [{\citenamefont {Lechner}\ \emph {et~al.}(2016)\citenamefont
  {Lechner}, \citenamefont {Maier}, \citenamefont {Hempel}, \citenamefont
  {Jurcevic}, \citenamefont {Lanyon}, \citenamefont {Monz}, \citenamefont
  {Brownnutt}, \citenamefont {Blatt},\ and\ \citenamefont
  {Roos}}]{Lechner2016}%
  \BibitemOpen
  \bibfield  {author} {\bibinfo {author} {\bibfnamefont {R.}~\bibnamefont
  {Lechner}}, \bibinfo {author} {\bibfnamefont {C.}~\bibnamefont {Maier}},
  \bibinfo {author} {\bibfnamefont {C.}~\bibnamefont {Hempel}}, \bibinfo
  {author} {\bibfnamefont {P.}~\bibnamefont {Jurcevic}}, \bibinfo {author}
  {\bibfnamefont {B.~P.}\ \bibnamefont {Lanyon}}, \bibinfo {author}
  {\bibfnamefont {T.}~\bibnamefont {Monz}}, \bibinfo {author} {\bibfnamefont
  {M.}~\bibnamefont {Brownnutt}}, \bibinfo {author} {\bibfnamefont
  {R.}~\bibnamefont {Blatt}},\ and\ \bibinfo {author} {\bibfnamefont {C.~F.}\
  \bibnamefont {Roos}},\ }\bibfield  {title} {\bibinfo {title}
  {Electromagnetically-induced-transparency ground-state cooling of long ion
  strings},\ }\href {https://doi.org/10.1103/PHYSREVA.93.053401} {\bibfield
  {journal} {\bibinfo  {journal} {Physical Review A}\ }\textbf {\bibinfo
  {volume} {93}},\ \bibinfo {pages} {053401} (\bibinfo {year}
  {2016})}\BibitemShut {NoStop}%
\bibitem [{\citenamefont {Häffner}\ \emph {et~al.}(2008)\citenamefont
  {Häffner}, \citenamefont {Roos},\ and\ \citenamefont {Blatt}}]{Haffner2008}%
  \BibitemOpen
  \bibfield  {author} {\bibinfo {author} {\bibfnamefont {H.}~\bibnamefont
  {Häffner}}, \bibinfo {author} {\bibfnamefont {C.~F.}\ \bibnamefont {Roos}},\
  and\ \bibinfo {author} {\bibfnamefont {R.}~\bibnamefont {Blatt}},\ }\bibfield
   {title} {\bibinfo {title} {Quantum computing with trapped ions},\ }\href
  {https://doi.org/10.1016/J.PHYSREP.2008.09.003} {\bibfield  {journal}
  {\bibinfo  {journal} {Physics Reports}\ }\textbf {\bibinfo {volume} {469}},\
  \bibinfo {pages} {155} (\bibinfo {year} {2008})}\BibitemShut {NoStop}%
\end{thebibliography}%

\end{document}